\pdfoutput=1
\RequirePackage{fixltx2e}
\documentclass[11pt,a4paper]{article}
\usepackage[utf8]{inputenc}
\usepackage{jheppub}
\hypersetup{unicode=true,bookmarksopen=true}

\usepackage{mathtools}
\usepackage{lmodern}
\usepackage[T1]{fontenc}
\usepackage{bbm}
\DeclareMathAlphabet{\mathbfi}{OML}{cmm}{b}{it}

\usepackage{subcaption}

\let\originalleft\left
\let\originalright\right
\renewcommand{\left}{\mathopen{}\mathclose\bgroup\originalleft}
\renewcommand{\right}{\aftergroup\egroup\originalright}

\makeatletter
\newenvironment{equations}[1][]{\subequations\ifx\relax#1\relax\else\label{#1}\fi\align\ignorespaces}{\endalign\ignorespacesafterend\endsubequations}
\def\@spliteq#1{\begin{equation}\begin{split}#1\end{split}\end{equation}}
\def\splitequation{\collect@body\@spliteq}

\makeatother

\renewcommand{\vec}[1]{{\ifnum9<1#1\mathbf{#1}\else\ifcat\noexpand#1\relax\boldsymbol{#1}\else\mathbfi{#1}\fi\fi}}

\newcommand{\mathi}{\mathrm{i}}
\let\oldre\Re
\let\oldim\Im
\renewcommand{\Re}{\oldre\mathfrak{e}\,}
\renewcommand{\Im}{\oldim\mathfrak{m}\,}
\newcommand{\total}{\mathop{}\!\mathrm{d}}

\newcommand{\unitmatrix}{\mathbbm{1}}
\newcommand{\tr}{\operatorname{tr}}
\newcommand{\eqend}[1]{\,#1}

\newcommand{\bra}[1]{\left\langle{#1}\right\vert}
\newcommand{\ket}[1]{\left\vert{#1}\right\rangle}

\newcommand{\normord}[1]{\mathopen{:}{#1}\mathclose{:}}

\allowdisplaybreaks

\begin{document}

\title{Trace anomaly for chiral fermions via Hadamard subtraction}

\author{Markus B. Fr{\"o}b and}
\author{Jochen Zahn}
\affiliation{Institut f{\"u}r Theoretische Physik, Universit{\"a}t Leipzig, Br{\"u}derstra{\ss}e 16, 04103 Leipzig, Germany}

\emailAdd{mfroeb@itp.uni-leipzig.de}
\emailAdd{jochen.zahn@itp.uni-leipzig.de}

\abstract{We calculate the trace (conformal) anomaly for chiral fermions in a general curved background using Hadamard subtraction. While in intermediate steps of the calculation imaginary terms proportional to the Pontryagin density appear, imposing a vanishing divergence of the stress tensor these terms completely cancel, and we recover the well-known result equal to half the trace anomaly of a Dirac fermion. We elaborate in detail on the advantages of the Hadamard method for the general definition of composite operators in general curved spacetimes, and speculate on possible causes for the appearance of the Pontryagin density in other calculations.}


\maketitle

\section{Introduction}

The trace anomaly for chiral fermions has recently received new attention through the work of Bonora et al.~\cite{bonoraetal2014}, who claim that an anomalous term proportional to the Pontryagin density $\mathcal{P} \equiv \frac{1}{2} \epsilon_{\alpha\beta\mu\nu} R^{\mu\nu\rho\sigma} R^{\alpha\beta}{}_{\rho\sigma}$ appears in the trace of the renormalised stress tensor operator $T^{\mu\nu}_\text{ren}$ (the possibility of such a term was first discussed by Nakayama~\cite{nakayama2012}, see also~\cite{nakayama2018}). Moreover, the coefficient of this term is imaginary, signaling a violation of unitarity since the Hamiltonian of the theory (which is the integral of the energy density $T^{00}_\text{ren}$ over space) becomes complex. The validity of this calculation has been put into doubt by Bastianelli and Martelli~\cite{bastianellimartelli2016}, who using Pauli--Villars regularisation and Fujikawa's method only recovered the standard result (half of the trace anomaly of a Dirac fermion), without any Pontryagin term. Another work by Bonora et al.~\cite{bonoraetal2017,bonoraetal2018} then pointed out some possible inconsistencies in the application of Pauli--Villars regularisation and Fujikawa's method to chiral theories, and rederived a non-vanishing anomalous term involving the Pontryagin density using dimensional regularisation.

Since also dimensional regularisation is not without problems when applied to theories with chiral fermions, we present here a derivation of the trace anomaly using Hadamard subtraction. In contrast to dimensional regularisation, Hadamard subtraction works directly in the physical dimension; in contrast to the Fujikawa method it does not introduce a mass term which either couples two fermions of different chirality or breaks Lorentz covariance. Moreover, Hadamard subtraction works directly in the physical Lorentzian spacetime and no continuation to Euclidean space is necessary, such that a formally Hermitean expression for a composite operator (such as the stress tensor, or a current) results in a renormalised operator which is also Hermitean. That a proper Hadamard subtraction of the stress tensor of a chiral fermion preserves Hermiticity was shown in~\cite{zahn2014b}. However, it was not shown that the resulting stress tensor is conserved (possibly after a further finite renormalisation). This gap is closed here, and the resulting trace anomaly is computed. We note that Hadamard subtraction does give the correct, undisputed axial anomaly for both Abelian and non-Abelian currents~\cite{zahn2014b}.

We will first revisit the definition of composite operators via Hadamard subtraction on the example of scalar fields in section~\ref{sec_hadamard}. This includes a rederivation of the well-known result for the scalar trace anomaly, emphasising the modern viewpoint on quantum field theory in curved spacetime~\cite{hollandswald2015}. In section~\ref{sec_chiral}, we then discuss the definition and renormalisation of the stress tensor of chiral fermions. We slightly deviate from the proposal of~\cite{zahn2014b}, in that we perform the Hadamard subtraction in a way that does not guarantee Hermiticity, noting that the possible violations of Hermiticity are local and covariant and can thus be absorbed in a further finite renormalisation. In a first step, we thus find an imaginary term, both in the divergence and the trace of the stress tensor. Performing the finite renormalisation that is necessary to obtain a conserved stress tensor also removes the imaginary Pontryagin term in the trace anomaly. We conclude with a discussion of our results and the comparison with other calculations.

We use the ``+++'' convention of~\cite{mtw1973}, and work in units with $\hbar = c = 1$.

\section{The definition of composite operators via Hadamard subtraction}
\label{sec_hadamard}

The class of quantum states for which Hadamard subtraction yields sensible results are the Hadamard states, whose $n$-point functions have a short-distance singularity structure of the same form as the vacuum in flat spacetime, with subleading corrections which are determined by the curvature~\cite{hadamard1932,baerginouxpfaeffle2007}. There is ample reason to consider Hadamard states as a physically distinguished class of states:
\begin{enumerate}
\item On a globally hyperbolic spacetime, Hadamard states always exist~\cite{fullingnarcowichwald1981}.
\item A state which is Hadamard in a neighborhood of a Cauchy surface will be Hadamard throughout~\cite{fullingsweenywald1978,sahlmannverch2001,sanders2010}.
\item Ground and thermal states in stationary spacetimes are Hadamard~\cite{fullingnarcowichwald1981,sahlmannverch2000}.
\item The conformal vacua for conformally coupled massless scalar fields and massless vector and spinor fields in conformally flat spacetimes (such as cosmological FLRW spacetimes) are Hadamard~\cite{parkertoms2009,pinamonti2009}.
\item For any mass or curvature coupling, the states of low energy on FLRW spacetimes~\cite{olbermann2007} and inhomogeneous expanding cosmological spacetimes~\cite{thembrum2013} are Hadamard.
\item Adiabatic vacuum states of infinite order are Hadamard~\cite{pirk1993,hollands2001,junkerschrohe2002}.
\end{enumerate}
Since an adiabatic state of adiabatic order $4$ is sufficient to obtain a finite renormalised stress tensor operator~\cite{parkertoms2009}, one might wonder whether the restriction to infinite adiabatic order is really necessary. For this, consider defining composite operators simply by Wick/normal ordering with respect to some state (instead of performing a Hadamard subtraction, which uses detailed information on the short-distance singularity structure of the Hadamard states). These composite operators thus have all vanishing expectation value in this state. However, if the state is not of Hadamard form, there exists a composite operator (for example the normal-ordered square of some time derivative of the scalar field) which has infinite fluctuations in this state~\cite{brunettifredenhagen2000,fewsterverch2013} --- clearly something which should be avoided on physical grounds. The class of physically reasonable states is thus seen to be the class of Hadamard states.

\subsection{Scalar fields}
\label{sec_hadamard_scalar}

For the scalar field in four spacetime dimensions, the two-point function of a Hadamard state $\ket{W}$ has the form~\cite{hadamard1932,baerginouxpfaeffle2007}
\begin{equation}
\label{wightman_4d_scalar2}
G^{+W}_{m^2}(x,x') \equiv - \mathi \bra{W} \phi(x) \phi(x') \ket{W} = H^+_{m^2}(x,x') - \frac{\mathi}{8 \pi^2} W_{m^2}(x,x')
\end{equation}
for all $x'$ in a normal geodesic neighbourhood of $x$, where $H^+_{m^2}$ is the Hadamard parametrix
\begin{equation}
\label{scalar_hadamard_parametrix}
H^+_{m^2}(x,x') \equiv - \frac{\mathi}{8 \pi^2} \left[ \frac{U_{m^2}}{\sigma_\epsilon} + V_{m^2} \ln\left( \mu^2 \sigma_\epsilon \right) \right]
\end{equation}
with the Wightman prescription $\sigma_\epsilon = \sigma + \mathi \epsilon (t-t')$. Here, $\sigma = \sigma(x,x')$ is Synge's world function~\cite{synge1960} which is equal to one half of the (signed) squared geodesic distance between $x$ and $x'$ and fulfils
\begin{equation}
\label{sigma_relations}
\nabla_\mu \sigma \nabla^\mu \sigma = 2 \sigma \eqend{,} \qquad \lim_{x' \to x} \sigma = 0 \eqend{,}
\end{equation}
$\mu$ is some scale used to make the argument of the logarithm dimensionless (and which for a massive field can be taken to be equal to the mass), and $U_{m^2}$, $V_{m^2}$ and $W_{m^2}$ are smooth symmetric biscalars depending on $m$. While the biscalars $U_{m^2}$ and $V_{m^2}$ and thus the Hadamard parametrix are completely determined by the local geometry (with the explicit formulas given below) and are thus state-independent, the biscalar $W_{m^2}$ depends on the quantum state.

For $x'$ close to $x$ (and thus small $\sigma$), one considers the expansion of the biscalars $U_{m^2}$, $V_{m^2}$ and $W_{m^2}$ of the form
\begin{equations}[hadamard_4d_scalarexpansion]
U_{m^2} &= U_{m^2}^{(0)} \eqend{,} \\
\{ V/W \}_{m^2} &= \sum_{k=0}^\infty \{ V/W \}_{m^2}^{(k)} \sigma^k
\end{equations}
with smooth biscalars $U_{m^2}^{(0)}$, $V_{m^2}^{(k)}$ and $W_{m^2}^{(k)}$, the former two usually referred to as Hadamard coefficients. While for a general spacetime this is only an asymptotic expansion, for analytic spacetimes it is even a convergent expansion (see, e.g., Refs.~\cite{moretti2000,hollandswald2005,decaninifolacci2008,baerginouxpfaeffle2007} and references therein). By requiring $G^{+W}_{m^2}$ to solve the Klein--Gordon equation
\begin{equation}
\label{scalar_eom_wightman}
\left( \nabla^2 - m^2 - \xi R \right) G^{+W}_{m^2}(x,x') = 0
\end{equation}
outside of coincidence and comparing manifest powers of $\sigma$, one obtains
\begin{equation}
\label{hadamard_4d_u0}
U_{m^2}^{(0)} = \sqrt{\Delta}
\end{equation}
with the van~Vleck--Morette determinant~\cite{dewittbrehme1960,synge1960} defined by either the explicit expression
\begin{equation}
\Delta(x,x') = - \left[ g(x) g(x') \right]^{-\frac{1}{2}} \det\left[ \nabla_\alpha \nabla_{\beta'} \sigma(x,x') \right]
\end{equation}
or the first-order differential equation
\begin{equation}
\label{delta_relations}
\nabla^\rho \sigma \nabla_\rho \ln \Delta = 4 - \nabla^2 \sigma \eqend{,} \qquad \lim_{x' \to x} \Delta = 1 \eqend{,}
\end{equation}
and the recursion relations
\begin{equations}[hadamard_4d_scalar_recursion]
Q_{2k+4} V_{m^2}^{(k+1)} &= - \frac{1}{k+1} \left( \nabla^2 - m^2 - \xi R \right) V_{m^2}^{(k)} \eqend{,} \\
Q_{2k+4} W_{m^2}^{(k+1)} &= - \frac{1}{k+1} \left[ \left( \nabla^2 - m^2 - \xi R \right) W_{m^2}^{(k)} + Q_{4k+6} V_{m^2}^{(k+1)} \right]
\end{equations}
with the first-order differential operator
\begin{equation}
\label{qk_def}
Q_k \equiv 2 \nabla^\mu \sigma \nabla_\mu - \nabla^\mu \sigma \nabla_\mu \ln \Delta + k \eqend{,}
\end{equation}
subject to the boundary condition
\begin{equation}
\label{hadamard_4d_v0_bdy}
Q_2 V_{m^2}^{(0)} = - \left( \nabla^2 - m^2 - \xi R \right) \sqrt{\Delta} \eqend{.}
\end{equation}
It is thus seen explicitly that $U_{m^2}$ and $V_{m^2}$ are completely determined geometrically, while for $W_{m^2}$ the first coefficient is an arbitrary symmetric solution of the free Klein--Gordon equation
\begin{equation}
\left( \nabla^2 - m^2 - \xi R \right) W_{m^2}^{(0)} = 0 \eqend{,}
\end{equation}
which encodes the state-dependence of the two-point function. Imposing smoothness, there is a unique solution to the recursion relations~\eqref{hadamard_4d_scalar_recursion} for which the coefficients are symmetric~\cite{moretti2000}. This solution can be given explicitly as an integral in Riemannian normal coordinates, but in the following we only need that the unique smooth solution to $Q_k f = 0$ is $f = 0$. We remark that the anti-symmetric part of the two-point function $G^{+W}_{m^2}(x,x')$ is uniquely fixed by the commutation relation
\begin{equation}
\label{hadamard_phi_commutator}
\left[ \phi(x), \phi(x') \right] = \mathi \left[ G_{m^2}^\text{adv}(x,x') - G_{m^2}^\text{ret}(x,x') \right] = \mathi \left[ H^+_{m^2}(x,x') - H^+_{m^2}(x',x) \right] \eqend{,}
\end{equation}
with $G_{m^2}^{\text{adv}/\text{ret}}(x,x')$ the uniquely defined advanced and retarded propagators.

The Feynman propagator has the same expansion
\begin{equation}
G^{\text{F},W}_{m^2}(x,x') \equiv - \mathi \bra{W} \mathcal{T} \phi(x) \phi(x') \ket{W} = H^\text{F}_{m^2}(x,x') - \frac{\mathi}{8 \pi^2} W_{m^2}(x,x') \eqend{,}
\end{equation}
where the parametrix $H^\text{F}_{m^2}$ is now given by~\eqref{scalar_hadamard_parametrix} with the Feynman prescription $\sigma_\epsilon = \sigma + \mathi \epsilon$, and fulfils
\begin{equation}
\left( \nabla^2 - m^2 - \xi R \right) G^{\text{F},W}_{m^2}(x,x') = \delta(x,x')
\end{equation}
with the covariant $\delta$ distribution
\begin{equation}
\delta(x,x') \equiv \frac{\delta^4(x-x')}{\sqrt{-g}} \eqend{.}
\end{equation}
Note that the biscalars $U_{m^2}$, $V_{m^2}$ and $W_{m^2}$ are always the same smooth functions, and that only the $\epsilon$ prescription needed to resolve the singularity at $\sigma = 0$ differs between the Wightman two-point function and the corresponding Feynman propagator.

Composite operators are now defined by point splitting and subtraction of singular terms~\cite{dirac1934,schwinger1951,dewitt1975,christensen1976,christensen1978}. Since the Hadamard parametrix contains all singular terms as $x' \to x$, one can define composite operators in two ways, for which we use the example of $\phi^2$:
\begin{itemize}
\item Wick/normal ordering with respect to the state $\ket{W}$:
\begin{equation}
\normord{\phi^2}_\text{W}(x) \equiv \lim_{x' \to x} \left[ \phi(x) \phi(x') - \mathi G^{+W}_{m^2}(x,x') \unitmatrix \right] \eqend{,}
\end{equation}
and we obtain a vanishing expectation value
\begin{equation}
\bra{W} \normord{\phi^2}_\text{W}(x) \ket{W} = 0 \eqend{.}
\end{equation}
\item Normal ordering with respect to the Hadamard parametrix only:
\begin{equation}
\normord{\phi^2}_\text{H}(x) \equiv \lim_{x' \to x} \left[ \phi(x) \phi(x') - \mathi H^+_{m^2}(x,x') \unitmatrix \right] \eqend{,}
\end{equation}
and we obtain the expectation value
\begin{equation}
\bra{W} \normord{\phi^2}_\text{H}(x) \ket{W} = \frac{1}{8 \pi^2} W_{m^2}(x,x) \eqend{,}
\end{equation}
which is finite because $W_{m^2}$ is a smooth function of both arguments. This prescription should be supplemented by a finite number of local renormalisation ambiguities, see below.
\end{itemize}
The second possibility is what properly defines the Hadamard subtraction method, and it has many advantages over the normal ordering with respect to a state. It was in fact already proposed by Dirac~\cite{dirac1934} in the context of Dirac fields in external potentials, and used in the calculation of the Uehling potential~\cite{uehling1935}.

Since it is impossible to choose a distinguished (``vacuum'') state which is Hadamard consistently for an arbitrary spacetime~\cite{fewsterverch2012,fewster2018}, the first possibility involves an arbitrary choice for each spacetime. In contrast, the second possibility is uniquely defined for an arbitrary spacetime. Moreover, since the Hadamard parametrix is covariantly constructed from the local geometry, the composite operators defined by Hadamard subtraction transform covariantly under coordinate changes~\cite{hollandswald2001,hollandswald2002} (Hadamard subtraction is a locally covariant renormalisation scheme). That is, the operator $\normord{\phi^2}_\text{H}(x)$ is indeed a scalar at $x$, just as the classical function $\phi^2(x)$ would be and as the notation suggests, while $\normord{\phi^2}_\text{W}(x)$ is not. Note that for both subtraction methods the differences in expectation values between two different states are uniquely defined and agree,
\begin{splitequation}
\bra{W'} \normord{\phi^2}_\text{W}(x) \ket{W'} - \bra{W} \normord{\phi^2}_\text{W}(x) \ket{W} &= \frac{1}{8 \pi^2} \left[ W'_{m^2}(x,x) - W_{m^2}(x,x) \right] \\
&= \bra{W'} \normord{\phi^2}_\text{H}(x) \ket{W'} - \bra{W} \normord{\phi^2}_\text{H}(x) \ket{W} \eqend{.}
\end{splitequation}

The last (and probably most important) point in favour of Hadamard subtraction concerns the uniqueness of the so-defined composite operators. Namely, Hollands and Wald~\cite{hollandswald2001,hollandswald2002,hollandswald2005} have shown that for any locally covariant scheme of defining composite operators and (their) time-ordered products, the ambiguity in their definition, i.e. the renormalisation freedom, is basically the same as in flat space. More concretely, for a monomial composite operator (a simple product of fields and their derivatives) containing $k$ fields, the freedom consists in adding a sum of coefficients times monomial composite operators containing at most $k-2$ fields, where the coefficients are polynomials in the curvature tensors, their covariant derivatives, and parameters appearing in the theory (such as the mass, or coupling constants). Moreover, they must be of the correct dimension. Since Hadamard subtraction is a locally covariant scheme as explained above, the only freedom for $\phi^2$ is thus given by
\begin{equation}
\normord{\phi^2}_\text{H}(x) \to \normord{\phi^2}_\text{H}(x) + \left[ c_1 m^2 + c_2 R(x) \right] \unitmatrix \eqend{,}
\end{equation}
where the numerical constants $c_1$ and $c_2$ may depend on the dimensionless coupling constants of the theory. This freedom accounts for different choices of the scale $\mu$ in \eqref{scalar_hadamard_parametrix}. It can also be used to fulfil other desirable properties, for example conservation of the stress tensor operator. In general, one must first find a basis of composite operators, from which all other composite operators are then obtained taking derivatives and linear combinations (see below for an explicit example in the case of the stress tensor). In contrast, for composite operators normal-ordered with respect to a state, the ambiguities contain in general arbitrary functions of spacetime instead of constants~\cite{brunettifredenhagen2000}.

\subsection{The stress tensor for free scalar fields}

In the following, we repeat the analysis of~\cite{hollandswald2005} of the stress tensor of a free scalar field in four spacetime dimensions. It is given by
\begin{splitequation}
T_{\mu\nu} &= \nabla_\mu \phi \nabla_\nu \phi - \frac{1}{2} g_{\mu\nu} \nabla^\rho \phi \nabla_\rho \phi - \frac{1}{2} g_{\mu\nu} m^2 \phi^2 - \xi \left( \nabla_\mu \nabla_\nu - g_{\mu\nu} \nabla^2 - R_{\mu\nu} + \frac{1}{2} g_{\mu\nu} R \right) \phi^2
\end{splitequation}
with the non-minimal coupling parameter $\xi$, and is a composite operator quadratic in the scalar field of engineering dimension $4$. A basis of composite operators at most quadratic in the scalar field and with engineering dimension $\leq 4$ is given by
\begin{equation}
\Phi^{(0)} \equiv \unitmatrix \eqend{,} \quad \Phi^{(1)} \equiv \phi \eqend{,} \quad \Phi^{(2)} \equiv \phi^2 \eqend{,} \quad \Phi^{(3)}_{\mu\nu} \equiv \nabla_\mu \phi \nabla_\nu \phi \eqend{.}
\end{equation}
All other composite operators with the above constraints can be written as linear combinations of these and their derivatives, e.g.,
\begin{equation}
\phi \nabla_\mu \phi = \frac{1}{2} \nabla_\mu \Phi^{(2)} \eqend{,} \quad \phi \nabla_\mu \nabla_\nu \phi = \frac{1}{2} \nabla_\mu \nabla_\nu \Phi^{(2)} - \Phi^{(3)}_{\mu\nu} \eqend{.}
\end{equation}
In terms of this basis, we obtain
\begin{equation}
T_{\mu\nu} = \left( \delta_\mu^\rho \delta_\nu^\sigma - \frac{1}{2} g_{\mu\nu} g^{\rho\sigma} \right) \Phi^{(3)}_{\rho\sigma} - \frac{1}{2} g_{\mu\nu} m^2 \Phi^{(2)} - \xi \left( \nabla_\mu \nabla_\nu - g_{\mu\nu} \nabla^2 - R_{\mu\nu} + \frac{1}{2} g_{\mu\nu} R \right) \Phi^{(2)} \eqend{,}
\end{equation}
and thus the renormalised stress tensor operator is given by
\begin{splitequation}
\label{scalar_tmunuren_def}
T_{\mu\nu}^\text{ren} &= \left( \delta_\mu^\rho \delta_\nu^\sigma - \frac{1}{2} g_{\mu\nu} g^{\rho\sigma} \right) \Phi^{(3),\text{ren}}_{\rho\sigma} - \frac{1}{2} g_{\mu\nu} m^2 \Phi^{(2),\text{ren}} \\
&\quad- \xi \left( \nabla_\mu \nabla_\nu - g_{\mu\nu} \nabla^2 - R_{\mu\nu} + \frac{1}{2} g_{\mu\nu} R \right) \Phi^{(2),\text{ren}} \eqend{.}
\end{splitequation}

The composite operators $\Phi^{(k),\text{ren}}$ are determined in any locally covariant renormalisation scheme (for example, by Hadamard subtraction), and their renormalisation freedom is therefore of the form
\begin{equations}[scalar_phibasis_redef]
\Phi^{(0),\text{ren}} &\to \Phi^{(0),\text{ren}} \eqend{,} \\
\Phi^{(1),\text{ren}} &\to \Phi^{(1),\text{ren}} \eqend{,} \\
\Phi^{(2),\text{ren}} &\to \Phi^{(2),\text{ren}} + \left[ c_1 m^2 + c_2 R \right] \Phi^{(0),\text{ren}} \eqend{,} \\
\Phi^{(3),\text{ren}}_{\mu\nu} &\to \Phi^{(3),\text{ren}}_{\mu\nu} + \left[ \sum_i c_{3,i} C^{(4,i)}_{\mu\nu} + \sum_i c_{4,i} m^2 C^{(2,i)}_{\mu\nu} + c_5 m^4 g_{\mu\nu} \right] \Phi^{(0),\text{ren}} \eqend{,}
\end{equations}
where the $c_k$ are numerical constants that may depend on the dimensionless coupling parameter $\xi$, and the $C^{(d,i)}_{\mu\nu}$ are combinations of curvature tensors and their derivatives of engineering dimension $d$ (e.g., $C^{(2,1)}_{\mu\nu} = R_{\mu\nu}$, $C^{(2,2)}_{\mu\nu} = g_{\mu\nu} R$). The renormalisation freedom in the renormalised stress tensor is thus given by
\begin{equation}
\label{scalar_stresstensor_redef}
T_{\mu\nu}^\text{ren} \to T_{\mu\nu}^\text{ren} + \delta T_{\mu\nu} \Phi^{(0)} = T_{\mu\nu}^\text{ren} + \delta T_{\mu\nu} \unitmatrix
\end{equation}
with
\begin{splitequation}
\label{scalar_stresstensor_deltatmunu}
\delta T_{\mu\nu} &= - \frac{1}{2} g_{\mu\nu} \left[ c_1 m^4 + 2 c_5 m^4 + c_2 m^2 R \right] + \xi \left( R_{\mu\nu} - \frac{1}{2} g_{\mu\nu} R \right) \left[ c_1 m^2 + c_2 R \right] \\
&\quad- c_2 \xi \left( \nabla_\mu \nabla_\nu R - g_{\mu\nu} \nabla^2 R \right) + \left( \delta_\mu^\rho \delta_\nu^\sigma - \frac{1}{2} g_{\mu\nu} g^{\rho\sigma} \right) \left[ \sum_i c_{3,i} C^{(4,i)}_{\rho\sigma} + \sum_i c_{4,i} m^2 C^{(2,i)}_{\rho\sigma} \right] \eqend{.}
\end{splitequation}

Using point splitting and Hadamard subtraction, we obtain explicitly
\begin{equations}[scalar_phibasis_hadamardren]
\Phi^{(0),\text{ren}}(x) &= \unitmatrix \eqend{,} \\
\Phi^{(1),\text{ren}}(x) &= \phi(x) \eqend{,} \\
\Phi^{(2),\text{ren}}(x) &= \normord{\phi^2}_\text{H}(x) = \lim_{x' \to x} \left[ \phi(x) \phi(x') - \mathi H^+_{m^2}(x,x') \unitmatrix \right] \eqend{,} \\
\Phi^{(3),\text{ren}}_{\mu\nu}(x) &= \normord{\nabla_\mu \phi \nabla_\nu \phi}_\text{H}(x) = \lim_{x' \to x} \nabla_\mu \nabla_{\nu'} \left[ \phi(x) \phi(x') - \mathi H^+_{m^2}(x,x') \unitmatrix \right] \eqend{,}
\end{equations}
where a primed derivative means that it acts at $x'$. We see that the unit operator and the basic field are not renormalised. For the expectation values, we obtain
\begin{equations}[scalar_phibases_expectation]
\bra{W} \Phi^{(0),\text{ren}}(x) \ket{W} &= 1 \eqend{,} \\
\bra{W} \Phi^{(1),\text{ren}}(x) \ket{W} &= 0 \eqend{,} \\
\bra{W} \Phi^{(2),\text{ren}}(x) \ket{W} &= \frac{1}{8 \pi^2} W_{m^2}(x,x) \eqend{,} \\
\bra{W} \Phi^{(3),\text{ren}}_{\mu\nu}(x) \ket{W} &= \frac{1}{8 \pi^2} \lim_{x' \to x} \nabla_\mu \nabla_{\nu'} W_{m^2}(x,x') \eqend{,}
\end{equations}
and since $W_{m^2}$ is a smooth function, they are finite. For the stress tensor expectation value, it follows that
\begin{splitequation}
\label{scalar_stresstensor_expectation}
\bra{W} T_{\mu\nu}^\text{ren} \ket{W} &= \left( \delta_\mu^\rho \delta_\nu^\sigma - \frac{1}{2} g_{\mu\nu} g^{\rho\sigma} \right) \bra{W} \Phi^{(3),\text{ren}}_{\rho\sigma} \ket{W} - \frac{1}{2} g_{\mu\nu} m^2 \bra{W} \Phi^{(2),\text{ren}} \ket{W} \\
&\quad- \xi \left( \nabla_\mu \nabla_\nu - g_{\mu\nu} \nabla^2 - R_{\mu\nu} + \frac{1}{2} g_{\mu\nu} R \right) \bra{W} \Phi^{(2),\text{ren}} \ket{W} \\
&= \frac{1}{8 \pi^2} \left( \delta_\mu^\rho \delta_\nu^\sigma - \frac{1}{2} g_{\mu\nu} g^{\rho\sigma} \right) \lim_{x' \to x} \nabla_\rho \nabla_{\sigma'} W_{m^2}(x,x') - \frac{m^2}{16 \pi^2} g_{\mu\nu} W_{m^2}(x,x) \\
&\quad - \frac{\xi}{8 \pi^2} \left( \nabla_\mu \nabla_\nu - g_{\mu\nu} \nabla^2 - R_{\mu\nu} + \frac{1}{2} g_{\mu\nu} R \right) W_{m^2}(x,x) \eqend{,}
\end{splitequation}
which again is finite. However, the renomalised stress tensor operator $T_{\mu\nu}^\text{ren}$ will in general neither be traceless nor conserved. For its divergence, we compute from equation~\eqref{scalar_tmunuren_def}
\begin{equation}
\nabla^\mu T_{\mu\nu}^\text{ren} = \nabla^\mu \Phi^{(3),\text{ren}}_{\mu\nu} - \frac{1}{2} g^{\rho\sigma} \nabla_\nu \Phi^{(3),\text{ren}}_{\rho\sigma} - \frac{1}{2} \left( m^2 + \xi R \right) \nabla_\nu \Phi^{(2),\text{ren}} \eqend{,}
\end{equation}
and its trace is given by
\begin{equation}
g^{\mu\nu} T_{\mu\nu}^\text{ren} = - g^{\mu\nu} \Phi^{(3),\text{ren}}_{\mu\nu} - 2 m^2 \Phi^{(2),\text{ren}} + \xi \left( 3 \nabla^2 - R \right) \Phi^{(2),\text{ren}} \eqend{.}
\end{equation}
Using Synge's rule~\cite{synge1960,poissonpoundvega2011}
\begin{equation}
\label{synge_rule}
\nabla_\mu \lim_{x' \to x} f(x,x') = \lim_{x' \to x} \left[ \nabla_\mu f(x,x') + \nabla_{\mu'} f(x,x') \right] \eqend{,}
\end{equation}
we calculate from equations~\eqref{scalar_phibasis_hadamardren}
\begin{splitequation}
\nabla^\mu \Phi^{(3),\text{ren}}_{\mu\nu} &= \nabla^\mu \lim_{x' \to x} \nabla_\mu \nabla_{\nu'} \left[ \phi(x) \phi(x') - \mathi H^+_{m^2}(x,x') \unitmatrix \right] \\
&= \lim_{x' \to x} \left( \nabla^2 + \nabla^{\mu'} \nabla_\mu \right) \nabla_{\nu'} \left[ \phi(x) \phi(x') - \mathi H^+_{m^2}(x,x') \unitmatrix \right]
\end{splitequation}
and
\begin{equation}
g^{\rho\sigma} \nabla_\nu \Phi^{(3),\text{ren}}_{\rho\sigma} = \lim_{x' \to x} \left( \nabla_\nu + \nabla_{\nu'} \right) \nabla^\rho \nabla_{\rho'} \left[ \phi(x) \phi(x') - \mathi H^+_{m^2}(x,x') \unitmatrix \right] \eqend{,}
\end{equation}
and thus
\begin{splitequation}
\nabla^\mu T_{\mu\nu}^\text{ren} &= \lim_{x' \to x} \left[ \nabla^2 \nabla_{\nu'} - \frac{1}{2} \left( m^2 + \xi R \right) \left( \nabla_\nu + \nabla_{\nu'} \right) + \frac{1}{2} \nabla^\mu \nabla_{\mu'} \nabla_{\nu'} - \frac{1}{2} \nabla_\nu \nabla^\mu \nabla_{\mu'} \right] \\
&\qquad\qquad\times \left[ \phi(x) \phi(x') - \mathi H^+_{m^2}(x,x') \unitmatrix \right] \\
&= \lim_{x' \to x} \left( \nabla^2 - m^2 - \xi R \right) \nabla_{\nu'} \left[ \phi(x) \phi(x') - \mathi H^+_{m^2}(x,x') \unitmatrix \right] \eqend{.}
\end{splitequation}
Here we used also that the point-split expression is symmetric, since its antisymmetric part vanishes by~\eqref{hadamard_phi_commutator}. We see that equation-of-motion terms remain, which is of course analogous to the classical result
\begin{equation}
\nabla^\mu T_{\mu\nu} = \left( \nabla^2 - m^2 - \xi R \right) \phi \nabla_\nu \phi \eqend{.}
\end{equation}
Since the field operator $\phi(x)$ does fulfil its equation of motion (for example, in a mode expansion), only the terms containing the Hadamard parametrix remain, and we obtain
\begin{equation}
\nabla^\mu T_{\mu\nu}^\text{ren} = - \mathi \lim_{x' \to x} \left( \nabla^2 - m^2 - \xi R \right) \nabla_{\nu'} H^+_{m^2}(x,x') \unitmatrix \eqend{.}
\end{equation}

Similarly, for the trace of the renormalised stress tensor operator we obtain
\begin{splitequation}
g^{\mu\nu} T_{\mu\nu}^\text{ren} &= \lim_{x' \to x} \Big[ - (1-6\xi) \nabla^\mu \nabla_{\mu'} - \xi (1-6\xi) R - 2 (1-3\xi) m^2 + 3 \xi \left( \nabla^2 - \xi R - m^2 \right) \\
&\qquad\qquad+ 3 \xi \left( \nabla^{\prime 2} - \xi R - m^2 \right) \Big] \left[ \phi(x) \phi(x') - \mathi H^+_{m^2}(x,x') \unitmatrix \right] \eqend{,}
\end{splitequation}
analogous to the classical result
\begin{splitequation}
g^{\mu\nu} T_{\mu\nu} &= - (1-6\xi) \nabla^\mu \phi \nabla_\mu \phi - \xi (1-6\xi) R \phi^2 - 2 (1-3\xi) m^2 \phi^2 \\
&\quad+ 6 \xi \phi \left( \nabla^2 - \xi R - m^2 \right) \phi \eqend{.}
\end{splitequation}
The trace of the renormalised stress tensor has thus two contributions: the regular one that appears when $\xi \neq \frac{1}{6}$ or $m \neq 0$, and which depends on the state (since it depends on the field operator $\phi$), and the anomalous one
\begin{equation}
\mathcal{A} \equiv - 6 \mathi \xi \lim_{x' \to x} \left( \nabla^2 - \xi R - m^2 \right) H^+_{m^2}(x,x') \unitmatrix \eqend{.}
\end{equation}
We see that both a possible divergence of the renormalised stress tensor operator and the anomalous contribution to its trace are proportional to the identity operator, and are thus state-independent. Whether one can remove one (or both) of them by the freedom in the definition of the composite operators~\eqref{scalar_phibasis_redef} depends on the explicit values of the derivatives of the Hadamard parametrix. It is well-known that~\cite{moretti2003,decaninifolacci2006}
\begin{equations}
\lim_{x' \to x} \left( \nabla^2 - m^2 - \xi R \right) H^+_{m^2}(x,x') &= - \frac{3 \mathi}{4 \pi^2} V^{(1)}_{m^2}(x,x) \eqend{,} \\
\lim_{x' \to x} \left( \nabla^2 - m^2 - \xi R \right) \nabla_{\nu'} H^+_{m^2}(x,x') &= - \frac{\mathi}{4 \pi^2} \nabla_\nu V^{(1)}_{m^2}(x,x) \eqend{,}
\end{equations}
and
\begin{splitequation}
V^{(1)}_{m^2}(x,x) = \frac{1}{1440} &\bigg[ 2 R^{\alpha\beta\gamma\delta} R_{\alpha\beta\gamma\delta} - 2 R^{\alpha\beta} R_{\alpha\beta} + 5 (1-6\xi)^2 R^2 + 12 (1-5\xi) \nabla^2 R \\
&\qquad- 60 m^2 (1-6\xi) R + 180 m^4 \bigg] \eqend{,}
\end{splitequation}
and thus
\begin{equations}
\nabla^\mu T_{\mu\nu}^\text{ren} &= - \frac{1}{4 \pi^2} \nabla_\nu V^{(1)}_{m^2}(x,x) \unitmatrix \eqend{,} \\
\mathcal{A} &= - \frac{9 \xi}{2 \pi^2} V^{(1)}_{m^2}(x,x) \unitmatrix \eqend{.}
\end{equations}

To obtain a renormalised stress tensor operator with vanishing divergence, we use the renormalisation freedom given by~\eqref{scalar_stresstensor_redef}. Namely, from equation~\eqref{scalar_stresstensor_deltatmunu} we calculate that
\begin{splitequation}
\label{scalar_stresstensor_divdeltamunu}
\nabla^\mu \delta T_{\mu\nu} &= - \frac{1}{2} c_2 m^2 \nabla_\nu R - \frac{1}{4} c_2 \xi \nabla_\nu R^2 \\
&\quad+ \left( \delta_\nu^\sigma \nabla^\rho - \frac{1}{2} g^{\rho\sigma} \nabla_\nu \right) \left[ \sum_i c_{3,i} C^{(4,i)}_{\rho\sigma} + \sum_i c_{4,i} m^2 C^{(2,i)}_{\rho\sigma} \right] \eqend{,}
\end{splitequation}
and by choosing
\begin{equations}
C^{(4,1)}_{\mu\nu} &= g_{\mu\nu} R^{\alpha\beta\gamma\delta} R_{\alpha\beta\gamma\delta} \eqend{,} &&\qquad C^{(4,2)}_{\mu\nu} = g_{\mu\nu} R^{\alpha\beta} R_{\alpha\beta} \eqend{,} \\
C^{(4,3)}_{\mu\nu} &= g_{\mu\nu} R^2 \eqend{,} &&\qquad C^{(4,4)}_{\mu\nu} = g_{\mu\nu} \nabla^2 R \eqend{,} \\
c_2 &= \frac{1-6\xi}{48 \pi^2} \eqend{,} &&\qquad c_{4,i} = c_1 = c_5 = 0 \eqend{,} \\
c_{3,1} &= - \frac{1}{2880 \pi^2} \eqend{,} &&\qquad c_{3,2} = \frac{1}{2880 \pi^2} \eqend{,} \\
c_{3,3} &= - \frac{1-6\xi}{1152 \pi^2} \eqend{,} &&\qquad c_{3,4} = - \frac{1-5\xi}{480 \pi^2} \eqend{,}
\end{equations}
we obtain
\begin{splitequation}
\delta T_{\mu\nu} &= - \frac{1-6\xi}{96 \pi^2} g_{\mu\nu} m^2 R + \frac{1-6\xi}{48 \pi^2} \xi \left( R R_{\mu\nu} - \nabla_\mu \nabla_\nu R \right) + \frac{1}{2880 \pi^2} g_{\mu\nu} R^{\alpha\beta\gamma\delta} R_{\alpha\beta\gamma\delta} \\
&\quad- \frac{1}{2880 \pi^2} g_{\mu\nu} R^{\alpha\beta} R_{\alpha\beta} + \frac{(1-6\xi) (1-12\xi)}{1152 \pi^2} g_{\mu\nu} R^2 + \frac{1+5\xi-60\xi^2}{480 \pi^2} g_{\mu\nu} \nabla^2 R \eqend{,}
\end{splitequation}
and the redefined renormalised stress tensor operator $\tilde{T}_{\mu\nu}^\text{ren} = T_{\mu\nu}^\text{ren} + \delta T_{\mu\nu} \unitmatrix$ has vanishing divergence. However, it still has a non-vanishing anomalous trace: taking for simplicity $m = 0$ and $\xi = \frac{1}{6}$ such that only the anomalous contribution to the trace remains, we obtain
\begin{splitequation}
\label{scalar_traceanom}
\tilde{\mathcal{A}} &= g^{\mu\nu} \tilde{T}_{\mu\nu}^\text{ren} = \left[ - \frac{3}{4 \pi^2} V^{(1)}_0(x,x) + g^{\mu\nu} \delta T_{\mu\nu} \right] \unitmatrix \\
&= \frac{1}{2880 \pi^2} \left( R^{\alpha\beta\gamma\delta} R_{\alpha\beta\gamma\delta} - R^{\alpha\beta} R_{\alpha\beta} + \nabla^2 R \right) \unitmatrix = \frac{1}{5760 \pi^2} \left( - \mathcal{E}_4 + 3 C^2 + 2 \nabla^2 R \right) \unitmatrix \eqend{,}
\end{splitequation}
where we rewrote the result using the four-dimensional Euler density $\mathcal{E}_4$ and the square of the Weyl tensor, given by
\begin{equations}[eulerweylsquare]
\mathcal{E}_4 &\equiv R^{\alpha\beta\gamma\delta} R_{\alpha\beta\gamma\delta} - 4 R^{\alpha\beta} R_{\alpha\beta} + R^2 \eqend{,} \\
C^2 &\equiv R^{\alpha\beta\gamma\delta} R_{\alpha\beta\gamma\delta} - 2 R^{\alpha\beta} R_{\alpha\beta} + \frac{1}{3} R^2 \eqend{.}
\end{equations}
To remove the anomalous trace, we try to perform another redefinition. In order not to introduce a non-vanishing divergence~\eqref{scalar_stresstensor_divdeltamunu}, we must take
\begin{equation}
c_2 = 0 \eqend{,} \qquad C^{(4,i)}_{\rho\sigma} = \left( \delta_\rho^\mu \delta_\sigma^\nu - \frac{1}{2} g_{\rho\sigma} g^{\mu\nu} \right) \tilde{C}^{(4,i)}_{\mu\nu}
\end{equation}
with conserved tensors $\tilde{C}^{(4,i)}_{\mu\nu}$: $\nabla^\mu \tilde{C}^{(4,i)}_{\mu\nu} = 0$. Because of the four-dimensional Gau{\ss}--Bonnet identity, there are only two independent tensors of this form, given by the traceless
\begin{splitequation}
\label{k1munu_def}
K^{(1)}_{\mu\nu} &\equiv \frac{1}{\sqrt{-g}} \frac{\delta}{\delta g^{\mu\nu}} \int C^2 \sqrt{-g} \total^4 x = - 4 \nabla^{(\alpha} \nabla^{\beta)} C_{\alpha\mu\beta\nu} - 2 R^{\alpha\beta} C_{\alpha\mu\beta\nu} \\
&= - 2 \nabla^2 R_{\mu\nu} + \frac{2}{3} \nabla_\mu \nabla_\nu R + \frac{1}{3} g_{\mu\nu} \nabla^2 R - 4 R_{\alpha\mu\beta\nu} R^{\alpha\beta} + g_{\mu\nu} R_{\alpha\beta} R^{\alpha\beta} + \frac{4}{3} R R_{\mu\nu} - \frac{1}{3} g_{\mu\nu} R^2
\end{splitequation}
and
\begin{equation}
\label{k2munu_def}
K^{(2)}_{\mu\nu} \equiv \frac{1}{\sqrt{-g}} \frac{\delta}{\delta g^{\mu\nu}} \int R^2 \sqrt{-g} \total^4 x = 2 \nabla_\mu \nabla_\nu R - 2 g_{\mu\nu} \nabla^2 R - 2 R R_{\mu\nu} + \frac{1}{2} g_{\mu\nu} R^2
\end{equation}
with
\begin{equation}
\label{k2munu_trace}
g^{\mu\nu} K^{(2)}_{\mu\nu} = - 6 \nabla^2 R \eqend{.}
\end{equation}
By choosing an appropriate multiple of $K^{(2)}$, we can therefore remove the term proportional to $\nabla^2 R$ from the anomalous trace~\eqref{scalar_traceanom}, but not the terms involving the Euler density or the square of the Weyl tensor --- which of course agrees with well-known results~\cite{christensen1978}.

Having recovered the well-known trace anomaly for scalar fields, we want to stress the following point: using Hadamard subtraction, we have first obtained a well-defined renormalised composite operator $T_{\mu\nu}^\text{ren}$~\eqref{scalar_tmunuren_def}, and only afterwards calculated its divergence and (anomalous) trace. The basis of renormalised composite operators on which it depends have a completely explicit form~\eqref{scalar_phibasis_hadamardren} with explicit results for the expectation values~\eqref{scalar_phibases_expectation} (and variances, ...) in any Hadamard state $\ket{W}$ (determined by its two-point function~\eqref{wightman_4d_scalar2}), and consequently the stress tensor operator has a completely explicit expectation value~\eqref{scalar_stresstensor_expectation}. That is, it is by no means necessary to first calculate the divergence or trace of the classical stress tensor and then regularise it via point-splitting and renormalise using Hadamard subtraction. Instead, Hadamard renormalisation gives a fully well-defined and finite renormalised stress tensor operator, whose divergence and trace can be straightforwardly computed. However, the computation can be done directly on the level of the renormalised field operator instead of on the level of expectation values or correlation functions by using the point-split form, showing thereby more clearly the state-independence of the anomaly.

\section{Chiral fermions}
\label{sec_chiral}

We now generalise the above constructions to chiral fermions. We consider left-handed Weyl fermions, which are Dirac fermions satisfying
\begin{equation}
\psi = \mathcal{P}_+ \psi = \frac{1}{2} \left( \unitmatrix + \gamma_* \right) \psi \eqend{,} \qquad \bar{\psi} = \bar{\psi} \mathcal{P}_- = \frac{1}{2} \bar{\psi} \left( \unitmatrix - \gamma_* \right) \eqend{.}
\end{equation}
The chiral matrix $\gamma_*$ is defined such that $\gamma_*^2 = \unitmatrix$, and the projectors $\mathcal{P}_\pm$ are thus idempotent, $\mathcal{P}_\pm^2 = \mathcal{P}_\pm$. The curved-space action for Weyl fermions reads
\begin{equation}
\label{weyl_action}
S = - \int \bar{\psi} \mathcal{P}_- \mathcal{D} \left[ \mathcal{P}_+ \psi \right] \sqrt{-g} \total^4 x = - \int \bar{\psi} \mathcal{P}_- \mathcal{D} \psi \sqrt{-g} \total^4 x
\end{equation}
with
\begin{equation}
\mathcal{D} \equiv \gamma^\mu \nabla_\mu \eqend{,} \qquad \overleftarrow{\mathcal{D}} \equiv \overleftarrow{\nabla}_\mu \gamma^\mu \eqend{,}
\end{equation}
and where we used that $\gamma^\mu \gamma_* = - \gamma_* \gamma^\mu$. The derivative appearing here is the spinor covariant one that includes the spin connection:
\begin{equation}
\label{spinor_covd}
\nabla_\mu \psi \equiv \partial_\mu \psi + \frac{1}{4} \omega_{\mu\rho\sigma} \gamma^{\rho\sigma} \psi \eqend{,}
\end{equation}
where the curved-space $\gamma$ matrices are obtained from the constant flat-space ones using the frame field (vierbein) $e_\mu{}^a$ as $\gamma^\mu \equiv g^{\mu\nu} e_\nu{}^a \eta_{ab} \gamma^b$. Higher-order $\gamma$ matrices are obtained by antisymmetrisation, $\gamma^{\mu_1 \cdots \mu_k} \equiv \gamma^{[\mu_1} \cdots \gamma^{\mu_k]}$, and the spin connection $\omega$ is determined from derivatives of the frame field:
\begin{equation}
\omega_{\mu\rho\sigma} = \omega_{\mu[\rho\sigma]} = \eta_{ab} \left( e_\sigma{}^a \partial_{[\mu} e_{\rho]}{}^b - e_\rho{}^a \partial_{[\mu} e_{\sigma]}{}^b + e_\mu{}^a \partial_{[\sigma} e_{\rho]}{}^b \right) \eqend{.}
\end{equation}

A variation of the frame field $e_\mu{}^a$ can be decomposed into its symmetric and antisymmetric part according to~\cite{forgerroemer2004}
\begin{splitequation}
\label{frame_var_sym_asym}
\delta e_\mu{}^a &= \frac{1}{2} \left( \delta e_\mu{}^a + \eta_{bc} g^{\nu\rho} e_\mu{}^b e_\nu{}^a \delta e_\rho{}^c \right) + \frac{1}{2} \left( \delta e_\mu{}^a - \eta_{bc} g^{\nu\rho} e_\mu{}^b e_\nu{}^a \delta e_\rho{}^c \right) \\
&= \frac{1}{2} g^{\nu\rho} e_\nu{}^a \delta g_{\mu\rho} + \Lambda_b{}^a e_\mu{}^b
\end{splitequation}
with
\begin{equation}
\Lambda_b{}^a \equiv \frac{1}{2} \eta_{bc} g^{\nu\rho} \left( e_\nu{}^c \delta e_\rho{}^a - e_\nu{}^a \delta e_\rho{}^c \right) \eqend{.}
\end{equation}
That is, it consists of a metric variation and a local Lorentz transformation. Under the local Lorentz transformation, spinors and $\gamma$ matrices transform, while only the (curved-space) $\gamma$ matrices transform under the metric variation (see~\cite{forgerroemer2004} for details). The action~\eqref{weyl_action} is invariant under a local Lorentz transformation, and the metric variation gives the stress tensor:
\begin{equation}
\label{weyl_stresstensor_vardef}
T^{\mu\nu} \equiv \frac{2}{\sqrt{-g}} \frac{\delta S}{\delta g_{\mu\nu}} \eqend{.}
\end{equation}

We compute the following useful metric variations:
\begin{equations}
\delta \sqrt{-g} &= \frac{1}{2} \sqrt{-g} g^{\mu\nu} \delta g_{\mu\nu} \eqend{,} \\
\delta \gamma^\mu &= - \frac{1}{2} g^{\mu\nu} \gamma^\rho \delta g_{\rho\nu} \eqend{,} \\
\delta \left( \omega_{\mu\rho\sigma} \gamma^{\rho\sigma} \right) &= \gamma^{\rho\sigma} \nabla_\sigma \delta g_{\rho\mu} \eqend{,}
\end{equations}
and thus
\begin{equation}
\delta \mathcal{D} \psi = \delta \left( \gamma^\mu \nabla_\mu \psi \right) = - \frac{1}{2} \gamma^\mu \nabla^\nu \psi \delta g_{\mu\nu} + \frac{1}{4} \gamma^\rho \psi g^{\mu\nu} \nabla_\rho \delta g_{\mu\nu} - \frac{1}{4} \gamma^\mu \psi \nabla^\nu \delta g_{\mu\nu} \eqend{,}
\end{equation}
where we also have used the $\gamma$ matrix product identities~\cite{kennedy1981}
\begin{equations}
\gamma^\mu \gamma^{\nu_1 \cdots \nu_n} &= \gamma^{\mu \nu_1 \cdots \nu_n} + \sum_{i=1}^n (-1)^{i+1} g^{\mu\nu_i} \gamma^{\nu_1 \cdots \nu_{i-1} \nu_{i+1} \cdots \nu_n} \eqend{,} \\
\gamma^{\nu_1 \cdots \nu_n} \gamma^\mu &= \gamma^{\nu_1 \cdots \nu_n \mu} + \sum_{i=1}^n (-1)^{i+n} g^{\mu\nu_i} \gamma^{\nu_1 \cdots \nu_{i-1} \nu_{i+1} \cdots \nu_n} \eqend{.}
\end{equations}
Performing the variation~\eqref{weyl_stresstensor_vardef}, one obtains~\cite{forgerroemer2004}
\begin{equation}
\label{weyl_stresstensor_def}
T^{\mu\nu} = \frac{1}{2} \bar{\psi} \gamma^{(\mu} \overleftrightarrow{\nabla}^{\nu)} \mathcal{P}_+ \psi + \frac{1}{2} g^{\mu\nu} \left[ \bar{\psi} \overleftarrow{\mathcal{D}} \mathcal{P}_+ \psi - \bar{\psi} \mathcal{P}_- \mathcal{D} \psi \right] \eqend{,}
\end{equation}
where $\overleftrightarrow{\nabla} = \nabla - \overleftarrow{\nabla}$. While usually the equation of motion-terms containing $\mathcal{D} \psi$ and $\bar{\psi} \overleftarrow{\mathcal{D}}$ are dropped, we have seen that such terms can give contributions to the renormalised stress tensor operator due to the Hadamard subtraction, and thus keep them.

\subsection{Hadamard expansion for spinors}
\label{sec_chiral_hadamard}

For spinor fields, the Hadamard expansion is more involved. Let us consider first massive Dirac fields. The Feynman propagator
\begin{equation}
G^\text{F}_m(x,x') \equiv - \mathi \bra{W} \mathcal{T} \psi(x) \bar{\psi}(x') \ket{W} \eqend{,}
\end{equation}
equal to the time-ordered two-point function and fulfilling
\begin{equation}
\label{eq:D_G_F}
\left( \mathcal{D} - m \right) G^\text{F}_m(x,x') = \delta(x,x') \unitmatrix = G^\text{F}_m(x,x') \left( - \overleftarrow{\mathcal{D}}' - m \right) \eqend{,}
\end{equation}
can be split into a geometrically determined part $H^\text{F}_m$ and a smooth part $W_m$ which is state dependent, similar to the scalar case:
\begin{equation}
G^\text{F}_m(x,x') = H^\text{F}_m(x,x') - \frac{\mathi}{8 \pi^2} W_m(x,x') \eqend{.}
\end{equation}
Using that
\begin{equation}
\label{spinor_hadamard_pmdef}
P_m \equiv \left( \mathcal{D} - m \right) \left( \mathcal{D} + m \right) = \nabla^2 - m^2 - \frac{1}{4} R
\end{equation}
has the form of a (spinorial) Klein--Gordon operator, the parametrix can be represented as
\begin{equation}
H_m(x,x') = \left( \mathcal{D} + m \right) \mathcal{H}_{m^2}(x,x') \eqend{,}
\end{equation}
with $\mathcal{H}_{m^2}$ the parametrix corresponding to the wave operator $P_m$, i.e., of the form
\begin{equation}
\label{spinor_hadamard_parametrix}
\mathcal{H}_{m^2}(x,x') \equiv - \frac{\mathi}{8 \pi^2} \left[ \frac{\mathcal{U}_{m^2}}{\sigma_\epsilon} + \mathcal{V}_{m^2} \ln\left( \mu^2 \sigma_\epsilon \right) \right] \eqend{,}
\end{equation}
with the same prescriptions as before (i.e., the Wightman prescription $\sigma_\epsilon = \sigma + \mathi \epsilon (t-t')$ for the two-point function and the Feynman prescription $\sigma_\epsilon = \sigma + \mathi \epsilon$ for the Feynman propagator), and the asymptotic expansions
\begin{equations}[hadamard_4d_spinorexpansion]
\mathcal{U}_{m^2} &= \mathcal{U}_{m^2}^{(0)} \eqend{,} \\
\mathcal{V}_{m^2} &= \sum_{k=0}^\infty \mathcal{V}_{m^2}^{(k)} \sigma^k
\end{equations}
with smooth biscalars $\mathcal{U}_{m^2}^{(0)}$ and $\mathcal{V}_{m^2}^{(k)}$. Analogously to the scalar case, it follows that
\begin{equation}
\label{hadamard_4d_calu0}
\mathcal{U}_{m^2}^{(0)} = \sqrt{\Delta} \, \mathcal{I}
\end{equation}
with the spinor parallel propagator $\mathcal{I}$ obtained as the unique solution of
\begin{equation}
\label{spinorparallel_relations}
\nabla^\mu \sigma \nabla_\mu \mathcal{I} = 0 \eqend{,} \qquad \lim_{x' \to x} \mathcal{I} = \unitmatrix \eqend{,}
\end{equation}
and, with the same operator $Q_k$~\eqref{qk_def} as before, the recursion relations
\begin{equations}[hadamard_4d_spinor_recursion]
Q_{2k+4} \mathcal{V}_{m^2}^{(k+1)} &= - \frac{1}{k+1} P_m \mathcal{V}_{m^2}^{(k)} \eqend{,} \\
Q_{2k+4} \mathcal{W}_{m^2}^{(k+1)} &= - \frac{1}{k+1} \left[ P_m \mathcal{W}_{m^2}^{(k)} + Q_{4k+6} \mathcal{V}_{m^2}^{(k+1)} \right]
\end{equations}
subject to the boundary condition
\begin{equation}
\label{hadamard_4d_calv0_bdy}
Q_2 \mathcal{V}_{m^2}^{(0)} = - P_m \sqrt{\Delta} \, \mathcal{I} \eqend{.}
\end{equation}
As in the scalar case, $\mathcal{U}_{m^2}$ and $\mathcal{V}_{m^2}$ are completely determined locally by these relations.

The time-ordered two-point function for chiral fermions $G_*$ is the same as for massless Dirac fermions, sandwiched between the appropriate chiral projectors. That is, for a Weyl fermion we simply have
\begin{equation}
G^\text{F}_*(x,x') = \mathcal{P}_+ G^\text{F}_0(x,x') \mathcal{P}_- \eqend{,}
\end{equation}
and similarly for the Wightman two-point function. The parametrix is obtained in the same way, i.e.,
\begin{equation}
\label{weylspinor_feynman}
G^\text{F}_*(x,x') = H^\text{F}_*(x,x') - \frac{\mathi}{8 \pi^2} W_*(x,x')
\end{equation}
with
\begin{equation}
\label{weylspinor_hadamard_parametrix}
H_*(x,x') \equiv \mathcal{P}_+ \mathcal{D} \mathcal{H}_0(x,x') \mathcal{P}_-
\end{equation}
and the appropriate Feynman/Wightman prescription for $\sigma_\epsilon$. To avoid confusion, we mention the well-known fact that the above time-ordered two-point function for left-handed Weyl fermions is only an inverse on the space of left-handed Weyl fermions and not on the space of Dirac fermions, and thus not a full propagator. Namely, we have
\begin{equations}
\mathcal{D} G^\text{F}_*(x,x') &= \delta(x,x') \mathcal{P}_- \eqend{,} \\
G^\text{F}_*(x,x') \overleftarrow{\mathcal{D}}' &= - \delta(x,x') \mathcal{P}_+ \eqend{,}
\end{equations}
as can easily be checked by direct computation, using the massless limit of equation~\eqref{eq:D_G_F}. However, for the calculation of the stress tensor operator it is only relevant that it is a two-point function; to remove all doubt we show in Appendix~\ref{appendix_2comp} that the same final result for the anomaly is also obtain with two-component fermions, for which the Feynman propagator is an inverse of the two-component Dirac operator.

Let us at this point remark upon an important property of the two-point function $G^{+W}_m(x,x')$. It is a spinorial bidistribution, and thus should be integrated against test (co-)spinors. Due to the Hermiticity of the expectation value $\bra{W} \cdot \ket{W}$, we then have
\begin{equation}
\left[ \iint \bar{\chi}(x) G^{+W}_m(x,x') \eta(x') \total x \total x' \right]^* = \iint \bar{\eta}(x) G^{+W}_m(x,x') \chi(x') \total x \total x'
\end{equation}
for all test spinors $\chi$ and $\eta$. It would of course be desirable to keep this property both for $H^+_m(x,x')$ and $W_m(x,x')$. In particular, when defining composite operators by point-splitting with respect to $H^+_m$, this would ensure that Hermitean expressions, such as the stress tensor, are defined as Hermitean operators. However, the definition~\eqref{weylspinor_hadamard_parametrix} of $H_*$ does not guarantee this. To ensure Hermiticity, one should instead choose the symmetrised prescription
\begin{equation}
\label{weylspinor_hadamard_symmetric}
H_*(x, x') = \frac{1}{2} \mathcal{P}_+ \left[ \mathcal{D} \mathcal{H}_0(x,x') - \mathcal{H}_0(x,x') \overleftarrow{\mathcal{D}}' \right] \mathcal{P}_- \eqend{,}
\end{equation}
as suggested in~\cite{zahn2014a,zahn2014b}. However, as shown in~\cite{zahn2014b} and also below, the difference between the parametrices~\eqref{weylspinor_hadamard_parametrix} and~\eqref{weylspinor_hadamard_symmetric} is smooth and determined locally by the geometric data, so we may as well work with~\eqref{weylspinor_hadamard_parametrix} and then perform the additional finite renormalisations that are necessary to achieve a conserved and Hermitean stress tensor. This will be our approach in the following.

\subsection{The stress tensor}
\label{sec_chiral_stress}

From the explicit expression~\eqref{weyl_stresstensor_def} for the stress tensor, we obtain
\begin{equation}
T^{\mu\nu} = \frac{1}{2} \Psi^{\mu\nu} - \frac{1}{2} g^{\mu\nu} g_{\rho\sigma} \Psi^{\rho\sigma}
\end{equation}
with
\begin{equation}
\Psi^{\mu\nu} \equiv \bar{\psi} \gamma^{(\mu} \overleftrightarrow{\nabla}^{\nu)} \mathcal{P}_+ \psi \eqend{.}
\end{equation}
We define the renormalised stress tensor operator $T^{\mu\nu}_\text{ren}$ thus as
\begin{equation}
T^{\mu\nu}_\text{ren} \equiv \frac{1}{2} \Psi^{\mu\nu}_\text{ren} - \frac{1}{2} g^{\mu\nu} g_{\rho\sigma} \Psi^{\rho\sigma}_\text{ren} \eqend{,}
\end{equation}
and the renormalised operator $\Psi^{\mu\nu}_\text{ren}$ by point splitting and Hadamard subtraction,
\begin{splitequation}
\label{psimunu_pointsplit_def}
\Psi^{\mu\nu}_\text{ren}(x) = \tr \lim_{x' \to x} &\bigg[ \mathcal{P}_+ \left( \psi(x) \bar{\psi}(x') - \mathi H_*(x,x') \right) \overleftarrow{\nabla}^{(\mu'} \gamma^{\nu')} \\
&\qquad- \gamma^{(\mu} \nabla^{\nu)} \left( \psi(x) \bar{\psi}(x') - \mathi H_*(x,x') \right) \mathcal{P}_- \bigg] \eqend{.}
\end{splitequation}
Here, $\tr$ denotes a trace in spinor space, and we have taken into account the minus sign from interchanging the two spinors. 

Let us stress again that this renormalised operator and by consequence the renormalised stress tensor operator has finite expectation value in any Hadamard state, namely
\begin{splitequation}
\bra{W} T^{\mu\nu}_\text{ren}(x) \ket{W} &= \frac{1}{2} \left( \delta^\mu_\alpha \delta^\nu_\beta - g^{\mu\nu} g_{\alpha\beta} \right) \bra{W} \Psi^{\alpha\beta}_\text{ren}(x) \ket{W} \\
&= \frac{1}{16 \pi^2} \left( \delta^\mu_\alpha \delta^\nu_\beta - g^{\mu\nu} g_{\alpha\beta} \right) \\
&\qquad\qquad\times \tr \lim_{x' \to x} \left[ \mathcal{P}_+ W_*(x,x') \overleftarrow{\nabla}^{(\alpha'} \gamma^{\beta')} - \gamma^{(\alpha} \nabla^{\beta)} W_*(x,x') \mathcal{P}_- \right] \eqend{,}
\end{splitequation}
where $W_*$ is the state-dependent part of the propagator~\eqref{weylspinor_feynman}.

In analogy to the scalar case, the renormalisation freedom is given by composite operators containing at least two fields less~\cite{zahn2014a}. Since the theory is massless and $\Psi^{\mu\nu}$ has engineering dimension $4$, the only freedom is thus
\begin{equation}
\label{psimunuren_freedom}
\Psi^{\mu\nu}_\text{ren} \to \Psi^{\mu\nu}_\text{ren} + C^{\mu\nu} \unitmatrix \eqend{,}
\end{equation}
where $C^{\mu\nu}$ is a symmetric tensor of dimension $4$ constructed from curvature tensors and their covariant derivatives. This entails the renormalisation freedom
\begin{equation}
\label{tmunuren_freedom}
T^{\mu\nu}_\text{ren} \to T^{\mu\nu}_\text{ren} + \frac{1}{2} \left( C^{\mu\nu} - g^{\mu\nu} g_{\rho\sigma} C^{\rho\sigma} \right) \unitmatrix
\end{equation}
for the renormalised stress tensor operator, which we will use to impose its covariant conservation. Calculating the divergence and the trace of the renormalised stress tensor operator, we obtain
\begin{equations}[weyl_stresstensor_divtrace]
\nabla_\mu T^{\mu\nu}_\text{ren} &= \frac{1}{2} \nabla_\mu \Psi^{\mu\nu}_\text{ren} - \frac{1}{2} g_{\rho\sigma} \nabla^\nu \Psi^{\rho\sigma}_\text{ren} \eqend{,} \\
g_{\mu\nu} T^{\mu\nu}_\text{ren} &= - \frac{3}{2} g_{\rho\sigma} \Psi^{\rho\sigma}_\text{ren} \eqend{.}
\end{equations}

To determine the divergence and trace of the renormalised $\Psi^{\mu\nu}_\text{ren}$, we could use again the point-split form~\eqref{psimunu_pointsplit_def} to show that possible anomalies are state-independent. However, for demonstration purposes we will work directly with the manifestly finite expectation value
\begin{equation}
\label{psimunu_expectation}
\bra{W} \Psi^{\alpha\beta}_\text{ren}(x) \ket{W} = \frac{1}{8 \pi^2} \tr \lim_{x' \to x} \left[ \mathcal{P}_+ W_*(x,x') \overleftarrow{\nabla}^{(\alpha'} \gamma^{\beta')} - \gamma^{(\alpha} \nabla^{\beta)} W_*(x,x') \mathcal{P}_- \right] \eqend{,}
\end{equation}
and state independence will follow if all terms involving $W_*$ disappear, being replaced by the Hadamard parametrix $H_*$. Using Synge's rule~\eqref{synge_rule}, the cyclicity of the trace, and commuting covariant derivatives it follows that
\begin{equations}
\begin{split}
\label{psimunu_expectation_divergence}
\nabla_\mu \bra{W} \Psi^{\mu\nu}_\text{ren}(x) \ket{W} &= \frac{1}{16 \pi^2} \tr \lim_{x' \to x} \bigg[ - \nabla^\nu \mathcal{D} W_*(x,x') \mathcal{P}_- - \gamma^\nu \mathcal{D}^2 W_*(x,x') \mathcal{P}_- \\
&\hspace{6em}+ \mathcal{D} W_*(x,x') \overleftarrow{\nabla}^{\nu'} \mathcal{P}_- - \mathcal{P}_+ \nabla^\nu W_*(x,x') \overleftarrow{\mathcal{D}}' \\
&\hspace{6em}+ \mathcal{P}_+ W_*(x,x') \overleftarrow{\mathcal{D}}' \overleftarrow{\nabla}^{\nu'} + \mathcal{P}_+ W_*(x,x') \overleftarrow{\mathcal{D}}'^2 \gamma^\nu \bigg] \eqend{,}
\end{split} \\
\label{psimunu_expectation_trace}
g_{\rho\sigma} \bra{W} \Psi^{\rho\sigma}_\text{ren}(x) \ket{W} &= \frac{1}{8 \pi^2} \tr \lim_{x' \to x} \bigg[ \mathcal{P}_+ W_*(x,x') \overleftarrow{\mathcal{D}}' - \mathcal{D} W_*(x,x') \mathcal{P}_- \bigg] \eqend{.}
\end{equations}
Since the Wightman two-point function satisfies the equations of motion
\begin{equation}
\mathcal{D} G^{+W}_0(x,x') = 0 = G^{+W}_0(x,x') \overleftarrow{\mathcal{D}}' \eqend{,}
\end{equation}
its state-dependent part satisfies
\begin{equation}
\mathcal{D} W_*(x,x') = - 8 \mathi \pi^2 \mathcal{D} H_*(x,x') \eqend{,} \qquad W_*(x,x') \overleftarrow{\mathcal{D}}' = - 8 \mathi \pi^2 \mathcal{D} H_*(x,x') \overleftarrow{\mathcal{D}}' \eqend{,}
\end{equation}
and the expectation values of divergence~\eqref{psimunu_expectation_divergence} and trace~\eqref{psimunu_expectation_trace} reduce to
\begin{equations}
\begin{split}
\label{psimunu_expectation_divergence_hadamard}
\nabla_\mu \bra{W} \Psi^{\mu\nu}_\text{ren}(x) \ket{W} &= \frac{\mathi}{2} \tr \lim_{x' \to x} \bigg[ \nabla^\nu \mathcal{D} H_*(x,x') \mathcal{P}_- + \gamma^\nu \mathcal{D}^2 H_*(x,x') \mathcal{P}_- \\
&\hspace{6em}- \mathcal{D} H_*(x,x') \overleftarrow{\nabla}^{\nu'} \mathcal{P}_- + \mathcal{P}_+ \nabla^\nu H_*(x,x') \overleftarrow{\mathcal{D}}' \\
&\hspace{6em}- \mathcal{P}_+ H_*(x,x') \overleftarrow{\mathcal{D}}' \overleftarrow{\nabla}^{\nu'} - \mathcal{P}_+ H_*(x,x') \overleftarrow{\mathcal{D}}'^2 \gamma^\nu \bigg] \eqend{,}
\end{split} \\
\label{psimunu_expectation_trace_hadamard}
g_{\rho\sigma} \bra{W} \Psi^{\rho\sigma}_\text{ren}(x) \ket{W} &= \mathi \tr \lim_{x' \to x} \bigg[ \mathcal{D} H_*(x,x') \mathcal{P}_- - \mathcal{P}_+ H_*(x,x') \overleftarrow{\mathcal{D}}' \bigg] \eqend{.}
\end{equations}
We see that both of them are again state-independent, and are determined geometrically from the parametrix alone. However, the determination of the actual coincidence values is somewhat more complicated than for the scalar field, due to the fact that the parametrix $H_*$ is determined~\eqref{weylspinor_hadamard_parametrix} by taking a derivative of another parametrix $\mathcal{H}_0$, and the recursion relations only hold for this second one~\eqref{hadamard_4d_spinor_recursion}.

\subsection{Coincidence limits for the spinor parametrix}

To determine the divergence~\eqref{psimunu_expectation_divergence_hadamard} and trace~\eqref{psimunu_expectation_trace_hadamard}, we need the following coincidence limits:
\begin{equations}
\mathcal{K}_1(x) &\equiv \lim_{x' \to x} \mathcal{D} H_*(x,x') \eqend{,} \\
\mathcal{K}_2(x) &\equiv \lim_{x' \to x} H_*(x,x') \overleftarrow{\mathcal{D}}' \eqend{,} \\
\mathcal{K}_3(x) &\equiv \lim_{x' \to x} \mathcal{D}^2 H_*(x,x') \eqend{,} \\
\mathcal{K}_4(x) &\equiv \lim_{x' \to x} H_*(x,x') \overleftarrow{\mathcal{D}}'^2 \eqend{,} \\
\mathcal{K}_5^\nu(x) &\equiv \lim_{x' \to x} \left[ \nabla^\nu \mathcal{D} H_*(x,x') - \mathcal{D} H_*(x,x') \overleftarrow{\nabla}'^\nu \right] \eqend{,} \\
\mathcal{K}_6^\nu(x) &\equiv \lim_{x' \to x} \left[ \nabla^\nu H_*(x,x') \overleftarrow{\mathcal{D}}' - H_*(x,x') \overleftarrow{\mathcal{D}}' \overleftarrow{\nabla}'^\nu \right] \eqend{,}
\end{equations}
which are all individually finite. The results can be inferred from~\cite{zahn2014b}, where expressions for general spacetime dimension in terms of Hadamard coefficients were derived. However, as different conventions where used there, and we would like to keep the present article self-contained, we rederive the results in the present conventions.

For the determination of the coincidence limits we use the transport equations of the Hadamard coefficients and the coincidence limits~\cite{poissonpoundvega2011}
\begin{equation}
\label{coincidence_geom_1}
\lim_{x' \to x} \nabla_\mu \sigma = 0 \eqend{,} \qquad \lim_{x' \to x} \nabla_\mu \nabla_\nu \sigma = g_{\mu\nu} \eqend{,} \qquad \lim_{x' \to x} \nabla_\mu \nabla_{\nu'} \sigma = - g_{\mu\nu} \eqend{,} \qquad \lim_{x' \to x} \nabla_\mu \Delta = 0 \eqend{.}
\end{equation}
We begin by calculating from equation~\eqref{weylspinor_hadamard_parametrix} that
\begin{splitequation}
H_*(x,x') &= - \frac{\mathi}{8 \pi^2} \mathcal{P}_+ \bigg[ - \frac{\nabla_\mu \sigma}{\sigma_\epsilon^2} \gamma^\mu \mathcal{U}_0^{(0)} + \frac{1}{\sigma_\epsilon} \left( \mathcal{D} \mathcal{U}_0^{(0)} + \nabla_\mu \sigma \sum_{k=0}^\infty \sigma^k \gamma^\mu \mathcal{V}_0^{(k)} \right) \\
&\hspace{6em}+ \ln\left( \mu^2 \sigma_\epsilon \right) \sum_{k=0}^\infty \sigma^k \left( \mathcal{D} \mathcal{V}_0^{(k)} + (k+1) \nabla_\mu \sigma \gamma^\mu \mathcal{V}_0^{(k+1)} \right) \bigg] \mathcal{P}_- \eqend{,}
\end{splitequation}
and then from this
\begin{splitequation}
\mathcal{D} H_*(x,x') &= - \frac{\mathi}{8 \pi^2} \mathcal{P}_- \bigg[ - \frac{1}{\sigma_\epsilon^2} Q_0 \mathcal{U}_0^{(0)} + \frac{1}{\sigma_\epsilon} \left( Q_2 \mathcal{V}_0^{(0)} + \mathcal{D}^2 \mathcal{U}_0^{(0)} \right) \\
&\hspace{6em}+ \sum_{k=0}^\infty \sigma^k \left[ Q_{2k+4} + 2 (k+1) \right] \mathcal{V}_0^{(k+1)} \\
&\hspace{6em}+ \ln\left( \mu^2 \sigma_\epsilon \right) \sum_{k=0}^\infty \sigma^k \left[ (k+1) Q_{2k+4} \mathcal{V}_0^{(k+1)} + \mathcal{D}^2 \mathcal{V}_0^{(k)} \right] \bigg] \mathcal{P}_- \eqend{,}
\end{splitequation}
with the operator $Q_k$ defined in equation~\eqref{qk_def}. Using the transport equations~\eqref{hadamard_4d_spinor_recursion} and boundary conditions~\eqref{hadamard_4d_calu0}, \eqref{hadamard_4d_calv0_bdy} with $m = 0$, it now follows that
\begin{equations}
Q_0 \mathcal{U}_0^{(0)} &= \left( 2 \nabla^\mu \sigma \nabla_\mu + \nabla^2 \sigma - 4 \right) \left( \sqrt{\Delta} \, \mathcal{I} \right) \nonumber \\
& = \left( \nabla^\mu \sigma \nabla_\mu \ln \Delta + \nabla^2 \sigma - 4 \right) \sqrt{\Delta} \, \mathcal{I} = 0 \eqend{,} \\
Q_2 \mathcal{V}_0^{(0)} &= - P_0 \sqrt{\Delta} \, \mathcal{I} = - \mathcal{D}^2 \mathcal{U}_0^{(0)} \eqend{,} \\
Q_{2k+4} \mathcal{V}_0^{(k+1)} &= - \frac{1}{k+1} P_0 \mathcal{V}_0^{(k)} = - \frac{1}{k+1} \mathcal{D}^2 \mathcal{V}_0^{(k)} \eqend{,}
\end{equations}
recalling that $P_0 = \nabla^2 - \frac{1}{4} R$~\eqref{spinor_hadamard_pmdef}, and thus
\begin{equation}
\mathcal{D} H_*(x,x') = - \frac{\mathi}{8 \pi^2} \mathcal{P}_- \left[ \sum_{k=0}^\infty \sigma^k Q_{4k+6} \mathcal{V}_0^{(k+1)} \right] \mathcal{P}_- \eqend{.}
\end{equation}
Taking the coincidence limit, all terms except the one with $k = 0$ vanish. Moreover, using~\eqref{coincidence_geom_1} it follows that
\begin{equation}
\lim_{x' \to x} Q_6 \mathcal{V}_0^{(1)} = 6 \mathcal{V}_0^{(1)} \eqend{,}
\end{equation}
so that we obtain
\begin{equation}
\mathcal{K}_1(x) = - \frac{3 \mathi}{4 \pi^2} \mathcal{P}_- \left[ \lim_{x' \to x} \mathcal{V}_0^{(1)} \right] \mathcal{P}_- \eqend{.}
\end{equation}
Taking another derivative and the subsequent coincidence limit results in
\begin{equation}
\lim_{x' \to x} \nabla_\mu \mathcal{D} H_*(x,x') = - \frac{\mathi}{\pi^2} \mathcal{P}_- \left[ \lim_{x' \to x} \nabla_\mu \mathcal{V}_0^{(1)} \right] \mathcal{P}_-
\end{equation}
and using also Synge's rule~\eqref{synge_rule} we obtain
\begin{equations}
\mathcal{K}_3(x) &= - \frac{\mathi}{\pi^2} \mathcal{P}_+ \gamma^\mu \left[ \lim_{x' \to x} \nabla_\mu \mathcal{V}_0^{(1)} \right] \mathcal{P}_- \eqend{,} \\
\begin{split}
\mathcal{K}_5^\nu(x) &= 2 \lim_{x' \to x} \left[ \nabla^\nu \mathcal{D} H_*(x,x') \right] - \nabla^\nu \lim_{x' \to x} \mathcal{D} H_*(x,x') \\
&= \frac{\mathi}{4 \pi^2} \mathcal{P}_- \left[ - 8 \lim_{x' \to x} \nabla^\nu \mathcal{V}_0^{(1)} + 3 \nabla^\nu \left( \lim_{x' \to x} \mathcal{V}_0^{(1)} \right) \right] \mathcal{P}_- \eqend{.}
\end{split}
\end{equations}

For the other three coincidence limits, we first need to determine $H_*(x,x') \overleftarrow{\mathcal{D}}'$. Note that since the coincidence limit of $H_*(x,x')$ itself does not exist, we cannot simply apply Synge's rule. Following~\cite{zahn2014b}, we consider first the sum
\begin{splitequation}
\label{spinor_coincidence_sum}
&\mathcal{D} \mathcal{H}_0(x,x') + \mathcal{H}_0(x,x') \overleftarrow{\mathcal{D}}' = \frac{\mathi}{8 \pi^2} \bigg[ \frac{1}{\sigma_\epsilon^2} \left( \nabla_\mu \sigma \gamma^\mu \mathcal{U}_0^{(0)} + \mathcal{U}_0^{(0)} \nabla_{\rho'} \sigma \gamma^{\rho'} \right) \\
&\qquad- \frac{1}{\sigma_\epsilon} \left( \mathcal{D} \mathcal{U}_0^{(0)} + \mathcal{U}_0^{(0)} \overleftarrow{\mathcal{D}}' + \nabla_\mu \sigma \gamma^\mu \mathcal{V}_0^{(0)} + \nabla_{\rho'} \sigma \mathcal{V}_0^{(0)} \gamma^{\rho'} \right) \\
&\qquad- \nabla_\mu \sigma \sum_{k=0}^\infty \sigma^k \gamma^\mu \mathcal{V}_0^{(k+1)} - \nabla_{\rho'} \sigma \sum_{k=0}^\infty \sigma^k \mathcal{V}_0^{(k+1)} \gamma^{\rho'} \\
&\qquad- \ln\left( \mu^2 \sigma_\epsilon \right) \sum_{k=0}^\infty \sigma^k \left( \mathcal{D} \mathcal{V}_0^{(k)} + \mathcal{V}_0^{(k)} \overleftarrow{\mathcal{D}}' + (k+1) \nabla_\mu \sigma \gamma^\mu \mathcal{V}_0^{(k+1)} + (k+1) \nabla_{\rho'} \sigma \mathcal{V}_0^{(k+1)} \gamma^{\rho'} \right) \bigg] \eqend{.}
\end{splitequation}
From the explicit expression~\eqref{hadamard_4d_calu0}, we obtain
\begin{equation}
\label{spinor_coincidence_singular}
\nabla_\mu \sigma \gamma^\mu \mathcal{U}_0^{(0)} + \mathcal{U}_0^{(0)} \nabla_{\rho'} \sigma \gamma^{\rho'} = \sqrt{\Delta} \left( \nabla_\mu \sigma \gamma^\mu \mathcal{I} + \mathcal{I} \gamma^{\rho'} \nabla_{\rho'} \sigma \right) = \sqrt{\Delta} \left( \nabla_\mu \sigma + g_\mu^{\rho'} \nabla_{\rho'} \sigma \right) \gamma^\mu \mathcal{I} = 0
\end{equation}
using that
\begin{equation}
\label{gamma_xxs_parallel}
\gamma^\nu g_\nu^{\beta'} \mathcal{I} = \mathcal{I} \gamma^{\beta'} \eqend{,} \qquad \gamma^{\beta'} g^\nu_{\beta'} \mathcal{I}^{-1} = \mathcal{I}^{-1} \gamma^\nu \eqend{,}
\end{equation}
with $g_\nu^{\mu'}$ the parallel transport of vectors, defined by
\begin{equation}
\label{parallel_relations}
\nabla^\mu \sigma \nabla_\mu g_\alpha{}^{\beta'} = 0 \eqend{,} \qquad \lim_{x' \to x} g_\alpha{}^{\beta'} = \delta_\alpha^\beta \eqend{.}
\end{equation}
In the last step, we used that the expression in parentheses in~\eqref{spinor_coincidence_singular} vanishes~\cite{poissonpoundvega2011}. Hence, the most singular term in~\eqref{spinor_coincidence_sum} vanishes.

To show that the other singular terms vanish as well, we follow~\cite{zahn2014b} and derive a transport equation with vanishing boundary term, and then use that the unique smooth solution to such an equation vanishes~\cite{moretti2000}. We thus calculate
\begin{splitequation}
T_U &\equiv \mathcal{D} \mathcal{U}_0^{(0)} + \mathcal{U}_0^{(0)} \overleftarrow{\mathcal{D}}' + \nabla_\mu \sigma \gamma^\mu \mathcal{V}_0^{(0)} + \nabla_{\rho'} \sigma \mathcal{V}_0^{(0)} \gamma^{\rho'} \\
&= \gamma^\mu \left( \frac{\nabla_\mu \Delta}{2 \sqrt{\Delta}} \mathcal{I} + \sqrt{\Delta} \, \nabla_\mu \mathcal{I} + \nabla_\mu \sigma \mathcal{V}_0^{(0)} \right) + \left( \frac{\nabla_{\rho'} \Delta}{2 \sqrt{\Delta}} \mathcal{I} + \sqrt{\Delta} \, \nabla_{\rho'} \mathcal{I} + \nabla_{\rho'} \sigma \mathcal{V}_0^{(0)} \right) \gamma^{\rho'} \eqend{,}
\end{splitequation}
and since all of $\nabla_\mu \Delta$, $\nabla_\mu \mathcal{I}$ and $\nabla_\mu \sigma$ vanish in the coincidence limit, we have the coincidence limit $\lim_{x' \to x} T_U = 0$. After a lengthy but straightforward calculation the corresponding transport equation can be derived as
\begin{splitequation}
Q_0 T_U &= \left( 2 \nabla^\nu \sigma \nabla_\nu - \nabla^\nu \sigma \nabla_\nu \ln \Delta \right) T_U \\
&= \mathcal{D} Q_0 \mathcal{U}_0^{(0)} + \left( Q_0 \mathcal{U}_0^{(0)} \right) \overleftarrow{\mathcal{D}}' + \nabla_\mu \sigma \gamma^\mu \left[ Q_2 \mathcal{V}_0^{(0)} + P_0 \mathcal{U}_0^{(0)} \right] \\
&\quad+ \nabla_{\rho'} \sigma \left[ Q_2 \mathcal{V}_0^{(0)} + P_0 \mathcal{U}_0^{(0)} \right] \gamma^{\rho'} - P_0 \left[ \nabla_{\rho'} \sigma \mathcal{U}_0^{(0)} \gamma^{\rho'} + \nabla_\mu \sigma \gamma^\mu \mathcal{U}_0^{(0)} \right] \\
&= 0 \eqend{,}
\end{splitequation}
which vanishes because of the transport equations fulfilled by $\mathcal{U}_0^{(0)}$ and $\mathcal{V}_0^{(0)}$ and the previous result. Therefore $T_U = 0$. Similarly, for the quantity
\begin{equation}
T_V^{(k)} \equiv \mathcal{D} \mathcal{V}_0^{(k)} + \mathcal{V}_0^{(k)} \overleftarrow{\mathcal{D}}' + (k+1) \nabla_\mu \sigma \gamma^\mu \mathcal{V}_0^{(k+1)} + (k+1) \nabla_{\rho'} \sigma \mathcal{V}_0^{(k+1)} \gamma^{\rho'}
\end{equation}
one calculates $\lim_{x' \to x} T_V^{(k)} = 0$ and the transport equation
\begin{splitequation}
Q_2 T_V^{(0)} &= \mathcal{D} \left[ Q_2 \mathcal{V}_0^{(0)} + P_0 \mathcal{U}_0^{(0)} \right] + \left[ Q_2 \mathcal{V}_0^{(0)} + P_0 \mathcal{U}_0^{(0)} \right] \overleftarrow{\mathcal{D}}' + \nabla_\mu \sigma \gamma^\mu \left[ Q_4 \mathcal{V}_0^{(1)} + P_0 \mathcal{V}_0^{(0)} \right] \\
&\quad+ \nabla_{\rho'} \sigma \left[ Q_4 \mathcal{V}_0^{(1)} + P_0 \mathcal{V}_0^{(0)} \right] \gamma^{\rho'} - P_0 T_U \eqend{,}
\end{splitequation}
and (for $k \geq 1$)
\begin{splitequation}
Q_{2k+2} T_V^{(k)} &= \mathcal{D} \left[ Q_{2k+2} \mathcal{V}_0^{(k)} + \frac{1}{k} P_0 \mathcal{V}_0^{(k-1)} \right] + \left[ Q_{2k+2} \mathcal{V}_0^{(k)} + \frac{1}{k} P_0 \mathcal{V}_0^{(k-1)} \right] \overleftarrow{\mathcal{D}}' \\
&\quad+ \nabla_\mu \sigma \gamma^\mu \left[ (k+1) Q_{2k+4} \mathcal{V}_0^{(k+1)} + P_0 \mathcal{V}_0^{(k)} \right] \\
&\quad+ \nabla_{\rho'} \sigma \left[ (k+1) Q_{2k+4} \mathcal{V}_0^{(k+1)} + P_0 \mathcal{V}_0^{(k)} \right] \gamma^{\rho'} - \frac{1}{k} P_0 T_V^{(k-1)} \eqend{.}
\end{splitequation}
All the terms in brackets vanish by the transport equations for the $\mathcal{V}_0^{(k)}$, and thus inductively at each order one obtains a homogeneous transport equation with vanishing boundary condition (in the coincidence limit), whose unique smooth solution vanishes. It follows that the $T_V^{(k)}$ vanish for all $k$, and we have
\begin{splitequation}
\label{eq:DH_0+H_0D_smooth}
\mathcal{D} \mathcal{H}_0(x,x') + \mathcal{H}_0(x,x') \overleftarrow{\mathcal{D}}' = - \frac{\mathi}{8 \pi^2} \left[ \nabla_\mu \sigma \sum_{k=0}^\infty \sigma^k \gamma^\mu \mathcal{V}_0^{(k+1)} + \nabla_{\rho'} \sigma \sum_{k=0}^\infty \sigma^k \mathcal{V}_0^{(k+1)} \gamma^{\rho'} \right] \eqend{.}
\end{splitequation}

From this result, we calculate
\begin{splitequation}
H_*(x,x') \overleftarrow{\mathcal{D}}' &= \mathcal{P}_+ \mathcal{D} \mathcal{H}_0(x,x') \overleftarrow{\mathcal{D}}' \mathcal{P}_+ \\
&= \mathcal{P}_+ \mathcal{D} \left[ \mathcal{D} \mathcal{H}_0(x,x') + \mathcal{H}_0(x,x') \overleftarrow{\mathcal{D}}' \right] \mathcal{P}_+ - \mathcal{P}_+ \mathcal{D}^2 \mathcal{H}_0(x,x') \mathcal{P}_+ \\
&= - \frac{\mathi}{8 \pi^2} \mathcal{P}_+ \mathcal{D} \left[ \nabla_\mu \sigma \sum_{k=0}^\infty \sigma^k \gamma^\mu \mathcal{V}_0^{(k+1)} + \nabla_{\rho'} \sigma \sum_{k=0}^\infty \sigma^k \mathcal{V}_0^{(k+1)} \gamma^{\rho'} \right] \mathcal{P}_+ \\
&\qquad+ \frac{\mathi}{8 \pi^2} \mathcal{P}_+ \left[ \sum_{k=0}^\infty \sigma^k Q_{4k+6} \mathcal{V}_0^{(k+1)} \right] \mathcal{P}_+
\end{splitequation}
and
\begin{splitequation}
&\mathcal{D} \left[ \nabla_\mu \sigma \sum_{k=0}^\infty \sigma^k \gamma^\mu \mathcal{V}_0^{(k+1)} + \nabla_{\rho'} \sigma \sum_{k=0}^\infty \sigma^k \mathcal{V}_0^{(k+1)} \gamma^{\rho'} \right] \\
&\quad= \sum_{k=0}^\infty \sigma^k \bigg[ \left( \nabla_\mu \sigma \gamma^\nu \gamma^\mu \nabla_\nu + \nabla^2 \sigma + 2 k \right) \mathcal{V}_0^{(k+1)} + \left( \nabla_{\rho'} \sigma \nabla_\nu + \nabla_\nu \nabla_{\rho'} \sigma \right) \gamma^\nu \mathcal{V}_0^{(k+1)} \gamma^{\rho'} \bigg] \\
&\qquad+ \nabla_\nu \sigma \nabla_{\rho'} \sigma \sum_{k=0}^\infty (k+1) \sigma^k \gamma^\nu \mathcal{V}_0^{(k+2)} \gamma^{\rho'} \eqend{.}
\end{splitequation}
In the coincidence limit, most terms again vanish and we obtain
\begin{equation}
\mathcal{K}_2(x) = \lim_{x' \to x} H_*(x,x') \overleftarrow{\mathcal{D}}' = \frac{\mathi}{8 \pi^2} \mathcal{P}_+ \lim_{x' \to x} \left[ 2 \mathcal{V}_0^{(1)} + \gamma_\mu \mathcal{V}_0^{(1)} \gamma^\mu \right] \mathcal{P}_+ \eqend{.}
\end{equation}

For the other coincidence limits, we have to act with more derivatives, and calculate
\begin{splitequation}
\nabla_\mu H_*(x,x') \overleftarrow{\mathcal{D}}' &= - \frac{\mathi}{8 \pi^2} \mathcal{P}_+ \nabla_\mu \bigg[ \sum_{k=0}^\infty \sigma^k \Big[ - \left( \nabla_\nu \sigma \gamma^\nu \mathcal{D} + 2 (k+1) \right) \mathcal{V}_0^{(k+1)} \\
&\hspace{12em}+ \left( \nabla_{\rho'} \sigma \mathcal{D} + \gamma^\nu \nabla_\nu \nabla_{\rho'} \sigma \right) \mathcal{V}_0^{(k+1)} \gamma^{\rho'} \Big] \\
&\hspace{6em}+ \nabla_\nu \sigma \nabla_{\rho'} \sigma \sum_{k=0}^\infty (k+1) \sigma^k \gamma^\nu \mathcal{V}_0^{(k+2)} \gamma^{\rho'} \bigg] \mathcal{P}_+ \\
&= \frac{\mathi}{8 \pi^2} \mathcal{P}_+ \sum_{k=0}^\infty \sigma^k \bigg[ \nabla_\mu \nabla_\nu \sigma \gamma^\nu \mathcal{D} \mathcal{V}_0^{(k+1)} + 2 (k+1) \nabla_\mu \mathcal{V}_0^{(k+1)} \\
&\qquad\qquad- \left( \nabla_\mu \nabla_{\rho'} \sigma \mathcal{D} + \gamma^\nu \nabla_\nu \nabla_{\rho'} \sigma \nabla_\mu + \gamma^\nu \nabla_\mu \nabla_\nu \nabla_{\rho'} \sigma \right) \mathcal{V}_0^{(k+1)} \gamma^{\rho'} \bigg] \mathcal{P}_+ \\
&\quad+ \text{ terms containing } \nabla \sigma \eqend{,} \\
\end{splitequation}
and thus, using the coincidence limits~\eqref{coincidence_geom_1},
\begin{splitequation}
&\lim_{x' \to x} \nabla_\mu H_*(x,x') \overleftarrow{\mathcal{D}}' \\
&\qquad= \frac{\mathi}{8 \pi^2} \mathcal{P}_+ \lim_{x' \to x} \left[ \gamma_\mu \gamma^\nu \nabla_\nu \mathcal{V}_0^{(1)} + 2 \nabla_\mu \mathcal{V}_0^{(1)} + \gamma_\nu \nabla_\mu \mathcal{V}_0^{(1)} \gamma^\nu + \gamma^\nu \nabla_\nu \mathcal{V}_0^{(1)} \gamma_\mu \right] \mathcal{P}_+ \eqend{.}
\end{splitequation}
Using also Synge's rule~\eqref{synge_rule}, it follows that
\begin{splitequation}
\mathcal{K}_4(x) &= \lim_{x' \to x} H_*(x,x') \overleftarrow{\mathcal{D}}'^2 = \left[ \nabla_\mu \lim_{x' \to x} \left( H_*(x,x') \overleftarrow{\mathcal{D}}' \right) - \lim_{x' \to x} \left( \nabla_\mu H_*(x,x') \overleftarrow{\mathcal{D}}' \right) \right] \gamma^\mu \\
&= \frac{\mathi}{8 \pi^2} \mathcal{P}_+ \bigg[ \nabla_\mu \lim_{x' \to x} \left[ 2 \mathcal{V}_0^{(1)} + \gamma_\nu \mathcal{V}_0^{(1)} \gamma^\nu \right] \gamma^\mu \\
&\qquad- \lim_{x' \to x} \left[ \gamma_\mu \gamma^\nu \nabla_\nu \mathcal{V}_0^{(1)} \gamma^\mu + \gamma_\nu \nabla_\mu \mathcal{V}_0^{(1)} \gamma^\nu \gamma^\mu + 2 \nabla_\mu \mathcal{V}_0^{(1)} \gamma^\mu + 4 \gamma^\nu \nabla_\nu \mathcal{V}_0^{(1)} \right] \bigg] \mathcal{P}_-
\end{splitequation}
and
\begin{splitequation}
\mathcal{K}_6^\nu(x) &= \lim_{x' \to x} \left[ \nabla^\nu H_*(x,x') \overleftarrow{\mathcal{D}}' - H_*(x,x') \overleftarrow{\mathcal{D}}' \overleftarrow{\nabla}^{\nu'} \right] \\
&= 2 \lim_{x' \to x} \left( \nabla^\nu H_*(x,x') \overleftarrow{\mathcal{D}}' \right) - \nabla^\nu \lim_{x' \to x} \left( H_*(x,x') \overleftarrow{\mathcal{D}}' \right) \\
&= \frac{\mathi}{8 \pi^2} \mathcal{P}_+ \bigg[ \lim_{x' \to x} \left[ 2 \gamma^\nu \gamma^\mu \nabla_\mu \mathcal{V}_0^{(1)} + 4 \nabla^\nu \mathcal{V}_0^{(1)} + 2 \gamma_\mu \nabla^\nu \mathcal{V}_0^{(1)} \gamma^\mu + 2 \gamma^\mu \nabla_\mu \mathcal{V}_0^{(1)} \gamma^\nu \right] \\
&\qquad- \nabla^\nu \lim_{x' \to x} \left[ 2 \mathcal{V}_0^{(1)} + \gamma_\mu \mathcal{V}_0^{(1)} \gamma^\mu \right] \bigg] \mathcal{P}_+ \eqend{.}
\end{splitequation}

Taking all together and using the cyclicity of the trace, we obtain for the divergence~\eqref{psimunu_expectation_divergence_hadamard} and trace~\eqref{psimunu_expectation_trace_hadamard} of $\Psi^{\mu\nu}_\text{ren}$ the following expressions:
\begin{splitequation}
\label{psimunu_expectation_divergence_coeffs}
&\nabla_\mu \bra{W} \Psi^{\mu\nu}_\text{ren}(x) \ket{W} = \frac{1}{16 \pi^2} \tr \bigg[ 2 \lim_{x' \to x} \nabla^\nu \mathcal{V}_0^{(1)} + 3 \nabla^\nu \lim_{x' \to x} \mathcal{V}_0^{(1)} - 2 \lim_{x' \to x} \nabla_\mu \mathcal{V}_0^{(1)} \gamma^{\mu\nu} \\
&\qquad+ \nabla_\mu \lim_{x' \to x} \mathcal{V}_0^{(1)} \gamma^{\mu\nu} - 6 \lim_{x' \to x} \nabla^\nu \mathcal{V}_0^{(1)} \gamma_* + \nabla^\nu \lim_{x' \to x} \mathcal{V}_0^{(1)} \gamma_* + \nabla_\mu \lim_{x' \to x} \mathcal{V}_0^{(1)} \gamma^{\mu\nu} \gamma_* \bigg]
\end{splitequation}
and
\begin{equation}
\label{psimunu_expectation_trace_coeffs}
g_{\rho\sigma} \bra{W} \Psi^{\rho\sigma}_\text{ren}(x) \ket{W} = \frac{1}{4 \pi^2} \tr \lim_{x' \to x} \left[ 3 \mathcal{V}_0^{(1)} - 2 \mathcal{V}_0^{(1)} \gamma_* \right] \eqend{.}
\end{equation}
Similarly to the scalar case, only the second coefficient $\mathcal{V}_0^{(1)}$ (and its derivatives) enter the result. To calculate their coincidence limits, we take the coincidence limit of the recursion relation~\eqref{hadamard_4d_spinor_recursion} and its first derivative, which results in
\begin{equations}
\lim_{x' \to x} \mathcal{V}_0^{(1)}(x,x') &= - \frac{1}{4} \lim_{x' \to x} P_0 \mathcal{V}_0^{(0)}(x,x') \eqend{,} \\
\lim_{x' \to x} \nabla_\mu \mathcal{V}_0^{(1)}(x,x') &= - \frac{1}{6} \lim_{x' \to x} \nabla_\mu P_0 \mathcal{V}_0^{(0)}(x,x') \eqend{,}
\end{equations}
where we used again the coincidence limits~\eqref{coincidence_geom_1}. We thus need the coincidence limits of $\mathcal{V}_0^{(0)}(x,x')$ and its derivatives up to third order, which we can obtain in the same way by taking the coincidence limit of equation~\eqref{hadamard_4d_calv0_bdy} and its derivatives. To evaluate them, we need coincidence limits of higher-order derivatives of all geometric objects: the world function $\sigma$, the van Vleck--Morette determinant $\Delta$, and the vector and spinor parallel transports $g_\mu{}^{\nu'}$ and $\mathcal{I}$. These can again be obtain recursively in the same way, taking the coincidence limit of the defining equations~\eqref{sigma_relations}, \eqref{delta_relations}, \eqref{parallel_relations}, \eqref{spinorparallel_relations} and their derivatives. The calculation is tedious by hand and best automated using a computer algebra system and tensor package such as xAct~\cite{xact}. In addition to the coincidence limits~\eqref{coincidence_geom_1}, one obtains in this way
\begin{equations}[coincidence_highorder]
\lim_{x' \to x} \nabla_{(\mu_1} \cdots \nabla_{\mu_k)} \sigma &= 0 \qquad (k \geq 3) \eqend{,} \\
\lim_{x' \to x} \nabla_\mu \nabla_\nu \Delta &= \frac{1}{3} R_{\mu\nu} \eqend{,} \\
\lim_{x' \to x} \nabla_\mu \nabla_\nu \nabla_\rho \Delta &= \frac{1}{2} \nabla_{(\mu} R_{\nu\rho)} \eqend{,} \\
\lim_{x' \to x} \nabla_{(\mu} \nabla_\nu \nabla_\rho \nabla_{\sigma)} \Delta &= \frac{1}{3} R_{(\mu\nu} R_{\rho\sigma)} + \frac{3}{5} \nabla_{(\mu} \nabla_\nu R_{\rho\sigma)} + \frac{2}{15} R_{\alpha(\mu\nu|\beta|} R^\alpha{}_{\rho\sigma)}{}^\beta \eqend{,} \\
\lim_{x' \to x} \nabla_{(\mu} \nabla_\nu \nabla_\rho \nabla_\sigma \nabla_{\tau)} \Delta &= \frac{5}{3} R_{(\mu\nu} \nabla_\rho R_{\sigma\tau)} + \frac{2}{3} \nabla_{(\mu} \nabla_\nu \nabla_\rho R_{\sigma\tau)} + \frac{2}{3} R_{\alpha(\mu\nu|\beta|} \nabla_\rho R^\alpha{}_{\sigma\tau)}{}^\beta \eqend{,} \\
\lim_{x' \to x} \nabla_{(\mu_1} \cdots \nabla_{\mu_k)} g_\alpha{}^{\beta'} &= 0 \qquad (k \geq 1) \eqend{,} \\
\lim_{x' \to x} \nabla_{(\mu_1} \cdots \nabla_{\mu_k)} \mathcal{I} &= 0 \qquad (k \geq 1) \eqend{.}
\end{equations}
Non-symmetrised derivatives can be easily obtained from these by commuting covariant derivatives, taking into account that for a spinor (and objects which transform as a spinor, like the spinor parallel transport $\mathcal{I}$) we have
\begin{equation}
\label{spinor_covd_commutator}
\left( \nabla_\mu \nabla_\nu - \nabla_\nu \nabla_\mu \right) \psi = \frac{1}{4} R_{\mu\nu\rho\sigma} \gamma^{\rho\sigma} \psi \eqend{.}
\end{equation}
By the above procedure, we then obtain
\begin{equations}[spinor_parametrix_coincidence]
\lim_{x' \to x} \mathcal{V}_0^{(0)}(x,x') &= \frac{1}{24} R \, \unitmatrix \eqend{,} \\
\lim_{x' \to x} \nabla_\mu \mathcal{V}_0^{(0)}(x,x') &= \frac{1}{24} \nabla_\alpha R_{\beta\mu} \, \gamma^{\alpha\beta} + \frac{1}{48} \nabla_\mu R \, \unitmatrix \eqend{,} \\
\begin{split}
\lim_{x' \to x} \mathcal{V}_0^{(1)}(x,x') &= - \frac{1}{17280} \left[ - 15 R^2 + 24 R_{\mu\nu} R^{\mu\nu} + 13 R_{\mu\nu\rho\sigma} R^{\mu\nu\rho\sigma} + 36 \nabla^2 R \right] \unitmatrix \\
&\quad+ \frac{1}{768} R_{\alpha\beta}{}^{\mu\nu} R_{\gamma\delta\mu\nu} \, \gamma^{\alpha\beta\gamma\delta} \eqend{,}
\end{split}
\end{equations}
and the very long expressions for $\lim_{x' \to x} \nabla_{(\mu} \nabla_{\nu)} \mathcal{V}_0^{(0)}(x,x')$, $\lim_{x' \to x} \nabla_{(\mu} \nabla_\nu \nabla_{\rho)} \mathcal{V}_0^{(0)}(x,x')$, and $\lim_{x' \to x} \nabla_\mu \mathcal{V}_0^{(1)}(x,x')$ are given in appendix~\ref{appendix_coincidence}. In simplifying the above, we have used the Bianchi identities for the Riemann tensor, and also the four-dimensional identity for the Weyl tensor~\cite{lovelock1970}
\begin{equation}
\label{weyl_4d_id}
C_{\mu}{}^{\rho\sigma\tau} C_{\nu\rho\sigma\tau} = \frac{1}{4} g_{\mu\nu} C^{\alpha\beta\gamma\delta} C_{\alpha\beta\gamma\delta} \eqend{.}
\end{equation}

\subsection{The renormalised stress tensor}

Finally, we can insert the coincidence limits~\eqref{spinor_parametrix_coincidence} into the expressions for the divergence~\eqref{psimunu_expectation_divergence_coeffs} and trace~\eqref{psimunu_expectation_trace_coeffs} and perform the trace in spinor space. Since the trace of any $\gamma$ matrix and their antisymmetrised products vanishes, most terms do not contribute. The only non-vanishing traces are
\begin{equation}
\label{gamma_trace}
\tr \unitmatrix = 4 \eqend{,} \qquad \tr \left( \gamma_* \gamma^{\mu\nu\rho\sigma} \right) = - 4 \mathi \epsilon^{\mu\nu\rho\sigma}
\end{equation}
with the completely antisymmetric tensor $\epsilon^{\mu\nu\rho\sigma}$, and it follows that
\begin{equations}
\lim_{x' \to x} \tr \mathcal{V}_0^{(1)} &= - \frac{1}{4320} \left[ 13 R^{\alpha\beta\gamma\delta} R_{\alpha\beta\gamma\delta} + 24 R^{\alpha\beta} R_{\alpha\beta} - 15 R^2 + 36 \nabla^2 R \right] \eqend{,} \\
\lim_{x' \to x} \tr \nabla_\mu \mathcal{V}_0^{(1)} &= - \frac{1}{2880} \nabla_\mu \left[ 7 R^{\alpha\beta\gamma\delta} R_{\alpha\beta\gamma\delta} + 8 R^{\alpha\beta} R_{\alpha\beta} - 5 R^2 + 12 \nabla^2 R \right] \eqend{,} \\
\lim_{x' \to x} \tr \left( \mathcal{V}_0^{(1)} \gamma_* \right) &= - \frac{\mathi}{96} R_{\alpha\beta\gamma\delta} (\star R)^{\alpha\beta\gamma\delta} \eqend{,} \\
\lim_{x' \to x} \tr \left( \nabla_\mu \mathcal{V}_0^{(1)} \gamma_* \right) &= - \frac{\mathi}{192} \nabla_\mu \left[ R_{\alpha\beta\gamma\delta} (\star R)^{\alpha\beta\gamma\delta} \right] \eqend{,} \\
\lim_{x' \to x} \tr \left( \mathcal{V}_0^{(1)} \gamma_{\rho\sigma} \right) &= 0 \eqend{,} \\
\lim_{x' \to x} \tr \left( \nabla^\rho \mathcal{V}_0^{(1)} \gamma_{\rho\sigma} \right) &= - \frac{1}{2880} \nabla_\sigma \left[ 7 R^{\alpha\beta\gamma\delta} R_{\alpha\beta\gamma\delta} + 8 R^{\alpha\beta} R_{\alpha\beta} - 5 R^2 + 12 \nabla^2 R \right] \eqend{,} \\
\lim_{x' \to x} \tr \left( \mathcal{V}_0^{(1)} \gamma_{\rho\sigma} \gamma_* \right) &= - \frac{\mathi}{48} \left[ (\star R)_{\sigma\alpha\beta\gamma} R_\rho{}^{\alpha\beta\gamma} - (\star R)_{\rho\alpha\beta\gamma} R_\sigma{}^{\alpha\beta\gamma} \right] = 0 \eqend{.}
\end{equations}
Here, the Hodge dual $\star$ is defined as
\begin{equation}
(\star R)_{\alpha\beta\gamma\delta} = \frac{1}{2} \epsilon_{\alpha\beta\mu\nu} R^{\mu\nu}{}_{\gamma\delta} \eqend{,}
\end{equation}
and in addition to the identity~\eqref{weyl_4d_id} we also used
\begin{equation}
\label{weyl_4d_id2}
(\star C)_{\rho\alpha}{}^{\gamma\delta} C^{\sigma\alpha}{}_{\gamma\delta} = \frac{1}{2} (\star C)_{\alpha\beta}{}^{\gamma\delta} C^{\alpha\beta}{}_{\gamma\delta} \delta_\rho^\sigma - (\star C)_{\gamma\delta}{}^{\sigma\alpha} C_{\rho\alpha}{}^{\gamma\delta} \eqend{,}
\end{equation}
which can be obtained from \cite[Thm. III.2]{edgarhoglund2002} by multiplying with the dual Weyl tensor.

From this, we obtain
\begin{splitequation}
\nabla_\mu \bra{W} \Psi^{\mu\nu}_\text{ren}(x) \ket{W} = - \frac{1}{11520 \pi^2} \nabla^\nu &\bigg[ 3 R^{\alpha\beta\gamma\delta} R_{\alpha\beta\gamma\delta} + 8 R^{\alpha\beta} R_{\alpha\beta} - 5 R^2 \\
&\qquad+ 12 \nabla^2 R - 15 \mathi R_{\alpha\beta\gamma\delta} (\star R)^{\alpha\beta\gamma\delta} \bigg]
\end{splitequation}
and
\begin{splitequation}
g_{\rho\sigma} \bra{W} \Psi^{\rho\sigma}_\text{ren}(x) \ket{W} = - \frac{1}{5760 \pi^2} &\bigg[ 13 R^{\alpha\beta\gamma\delta} R_{\alpha\beta\gamma\delta} + 24 R^{\alpha\beta} R_{\alpha\beta} - 15 R^2 \\
&\qquad+ 36 \nabla^2 R - 30 \mathi R_{\alpha\beta\gamma\delta} (\star R)^{\alpha\beta\gamma\delta} \bigg] \eqend{,}
\end{splitequation}
and from this by equation~\eqref{weyl_stresstensor_divtrace}
\begin{equations}[weyl_stresstensor_expectationdivtrace]
\begin{split}
\nabla_\mu \bra{W} T^{\mu\nu}_\text{ren} \ket{W} &= \frac{1}{23040 \pi^2} \nabla^\nu \bigg[ 23 R^{\alpha\beta\gamma\delta} R_{\alpha\beta\gamma\delta} + 40 R^{\alpha\beta} R_{\alpha\beta} - 25 R^2 \\
&\hspace{8em}+ 60 \nabla^2 R - 45 \mathi R_{\alpha\beta\gamma\delta} (\star R)^{\alpha\beta\gamma\delta} \bigg] \eqend{,}
\end{split} \\
\begin{split}
g_{\mu\nu} \bra{W} T^{\mu\nu}_\text{ren} \ket{W} &= \frac{1}{3840 \pi^2} \bigg[ 13 R^{\alpha\beta\gamma\delta} R_{\alpha\beta\gamma\delta} + 24 R^{\alpha\beta} R_{\alpha\beta} - 15 R^2 \\
&\hspace{8em}+ 36 \nabla^2 R - 30 \mathi R_{\alpha\beta\gamma\delta} (\star R)^{\alpha\beta\gamma\delta} \bigg] \eqend{.}
\end{split}
\end{equations}
We see that terms proportional to the Pontryagin density $R_{\alpha\beta\gamma\delta} (\star R)^{\alpha\beta\gamma\delta}$ appear with an imaginary coefficient, similar to the results of Bonora et al.~\cite{bonoraetal2014,bonoraetal2017,bonoraetal2018}. These terms arise whenever four $\gamma$ matrices appearing in the coincidence limit of the spinor parametrix are traced together with one $\gamma_*$ coming from the chiral projectors, according to equation~\eqref{gamma_trace}. The appearance of imaginary terms is a consequence of defining the parametrix $H_*$ in a non-symmetric way (see the discussion at the end of section~\ref{sec_chiral_hadamard}). However, according to equation~\eqref{eq:DH_0+H_0D_smooth} our prescription~\eqref{weylspinor_hadamard_parametrix} and the proper symmetric prescription~\eqref{weylspinor_hadamard_symmetric} differ only by local geometric terms, and both renormalisation schemes are thus locally covariant. As explained in section~\ref{sec_hadamard_scalar} for scalar fields, and later generalised to Dirac spinors in~\cite{dappiaggihackpinamonti2009,zahn2014a,zahn2014b} (see section~\ref{sec_chiral_stress}), renormalised operators defined using two different locally covariant renormalisation schemes are related by the usual renormalisation freedom, which in this case is~\eqref{psimunuren_freedom}
\begin{equation}
\Psi^{\mu\nu}_\text{ren} \to \Psi^{\mu\nu}_\text{ren} + C^{\mu\nu} \unitmatrix \eqend{,}
\end{equation}
where $C^{\mu\nu}$ is a symmetric tensor of dimension $4$ constructed from curvature tensors and their covariant derivatives. We will see that if we use this freedom to obtain a conserved stress tensor, we automatically also remove the imaginary term proportional to the Pontryagin density from the trace anomaly.

The change~\eqref{psimunuren_freedom} entails the change~\eqref{tmunuren_freedom} in the stress tensor
\begin{equation}
T^{\mu\nu}_\text{ren} \to \tilde{T}^{\mu\nu}_\text{ren} = T^{\mu\nu}_\text{ren} + \frac{1}{2} \left( C^{\mu\nu} - g^{\mu\nu} g_{\rho\sigma} C^{\rho\sigma} \right) \unitmatrix \eqend{,}
\end{equation}
and by taking
\begin{equation}
\label{cmunu_redef}
C^{\mu\nu} = \frac{1}{34560 \pi^2} g^{\mu\nu} \left[ 23 R^{\alpha\beta\gamma\delta} R_{\alpha\beta\gamma\delta} + 40 R^{\alpha\beta} R_{\alpha\beta} - 25 R^2 + 60 \nabla^2 R - 45 \mathi R_{\alpha\beta\gamma\delta} (\star R)^{\alpha\beta\gamma\delta} \right] \eqend{,}
\end{equation}
the divergence and trace of the expectation value~\eqref{weyl_stresstensor_expectationdivtrace} of the changed stress tensor read
\begin{equations}[weyl_stresstensor_expectationdivtrace_new]
\nabla_\mu \bra{W} \tilde{T}^{\mu\nu}_\text{ren} \ket{W} &= 0 \eqend{,} \\
g_{\mu\nu} \bra{W} \tilde{T}^{\mu\nu}_\text{ren} \ket{W} &= \frac{1}{11520 \pi^2} \left[ - 7 R^{\alpha\beta\gamma\delta} R_{\alpha\beta\gamma\delta} - 8 R^{\alpha\beta} R_{\alpha\beta} + 5 R^2 - 12 \nabla^2 R \right] \eqend{.}
\end{equations}
We note that this finite renormalisation has not only removed the parity-odd Pontryagin density from the trace, but has also changed the coefficients of the parity-even terms. Similar to the scalar case, we can now try to perform another redefinition to also remove the anomalous trace. In order not to introduce a non-vanishing divergence, we must take
\begin{equation}
C^{\mu\nu} = \left( g^{\mu\rho} g^{\nu\sigma} - \frac{1}{3} g^{\mu\nu} g^{\rho\sigma} \right) \left( \alpha_1 K^{(1)}_{\rho\sigma} + \alpha_2 K^{(2)}_{\rho\sigma} \right)
\end{equation}
with the two independent conserved tensors of dimension four $K^{(i)}_{\rho\sigma}$ given in equations~\eqref{k1munu_def} and~\eqref{k2munu_def}. Since $K^{(1)}_{\mu\nu}$ is traceless, while~\eqref{k2munu_trace} $g^{\mu\nu} K^{(2)}_{\mu\nu} = - 6 \nabla^2 R$, this entails the change
\begin{equation}
\frac{\alpha_2}{2} g^{\rho\sigma} K^{(2)}_{\rho\sigma} = - 3 \alpha_2 \nabla^2 R
\end{equation}
in the trace of the renormalised stress tensor. Taking $\alpha_2 = - 1/(2880 \pi^2)$, we can remove the term proportional to $\nabla^2 R$ and obtain
\begin{splitequation}
\label{tmunu_traceanom_final}
g_{\mu\nu} \bra{W} \tilde{T}^{\mu\nu}_\text{ren} \ket{W} &= \frac{1}{11520 \pi^2} \left[ - 7 R^{\alpha\beta\gamma\delta} R_{\alpha\beta\gamma\delta} - 8 R^{\alpha\beta} R_{\alpha\beta} + 5 R^2 \right] \\
&= \frac{1}{11520 \pi^2} \left( 11 \mathcal{E}_4 - 18 C^2 \right) \eqend{,}
\end{splitequation}
with the Euler density $\mathcal{E}_4$ and the square of the Weyl tensor $C^2$ given in equation~\eqref{eulerweylsquare}. This is exactly half of the result for a Dirac spinor (see for example Refs.~\cite{christensen1978,dappiaggihackpinamonti2009,bastianellimartelli2016}).

\section{Discussion}
\label{sec_discussion}

We have calculated the trace anomaly for chiral fermions using the Hadamard subtraction method. Imposing conservation of the renormalised stress tensor operator, no imaginary terms proportional to the Pontryagin density remains in the trace anomaly, and the result is half of the trace anomaly for a Dirac fermion. This is in agreement with the results of Bastianelli and Martelli~\cite{bastianellimartelli2016} using Pauli--Villars regularisation, but does not agree with the work of Bonora et al.~\cite{bonoraetal2014,bonoraetal2017,bonoraetal2018}. Since it has been shown~\cite{hollandswald2001,hollandswald2002,hollandswald2005,dappiaggihackpinamonti2009,zahn2014a,zahn2014b} that any locally covariant renormalisation scheme (i.e., where the renormalised composite operators transform covariantly under coordinate changes, or Lorentz transformations in flat space) gives the same result up to the allowed finite renormalisation freedom, we thus have to look more closely into the derivation of the trace anomaly by Bonora et al. In their first article~\cite{bonoraetal2014}, they first derive the trace anomaly from a heat kernel calculation in Euclidean space, based on old results by Christensen and Duff~\cite{christensenduff1979}. However, while for bosons the Wick rotation from Minkowskian to Euclidean spacetime that is needed to apply these results is quite straightforward, the continuation of fermions is more subtle. To our knowledge, there seem to be two consistent possibilities:
\begin{enumerate}
\item The continuous Wick rotation derived by Mehta~\cite{mehta1990} and van Nieuwenhuizen and Waldron~\cite{vannieuwenhuizenwaldron1996}: here fields are transformed by multiplication with a rotation matrix depending on an angle $\theta$, such that for $\theta = 0$ the Minkowskian theory results, while for $\theta = \pi/2$ the Euclidean theory is obtained. For spinors, in order to obtain an action that is invariant under the Euclidean rotation group $\mathrm{SO}(4)$ from a Minkowskian action invariant under the Lorentz group $\mathrm{SO}(3,1)$, it is necessary to perform this transformation separately for spinors and cospinors, and in Euclidean space they are therefore independent and unrelated by Hermitean or complex conjugation. For chiral spinors in particular, while a left-handed spinor in Minkowski spacetime is transformed into a left-handed spinor in Euclidean space, a left-handed cospinor is transformed into a right-handed cospinor. The resulting Euclidean action thus couples left-handed and right-handed spinors, and the corresponding stress tensor does as well. In this approach, the action stays invariant under (Euclidean) chiral transformations, and one can define the corresponding anomaly. However, since the stress tensor does involve both right- and left-handed Weyl spinors, and in a symmetric way, it seems to us that using the heat kernel method one should take the average of the heat kernel coefficients for the squared Dirac operator acting on left- and right-handed spinors. In this way, the coefficicient of the Pontryagin density cancels, and one obtains a result consistent with ours (and other methods).
\item The analytic continuation of the vielbein while keeping the fields fixed as explained by Wetterich~\cite{wetterich2011}: here $\gamma$ matrices and spinors are unchanged, and consequently the Euclidean action couples left-handed cospinors with left-handed spinors. However, the resulting spinors only transform properly under Euclidean rotations if a different complex structure is used, which is not compatible with chiral invariance. (One could also use the Minkowskian complex structure, which then preserves invariance under chiral transformations but is not compatible with Euclidean rotations, see also~\cite{kupschthacker1990}.) That is, within this approach one has to choose between Lorentz invariance and invariance under chiral transformations, and both cannot be realised simultaneously. It does not seem to us that this approach is very suitable for the calculation of a Minkowski trace anomaly, where the classical stress tensor is both invariant under chiral transformations and transforms properly under a Lorentz transformation, even though it certainly can be done in some way.
\end{enumerate}

Bonora et al.\ then support their result by an explicit perturbative calculation using dimensional regularisation, clarifying some steps of the calculation in their other articles~\cite{bonoraetal2017,bonoraetal2018}. It is well-known that the definition of the chiral matrix $\gamma_*$ in dimensional regularisation is non-trivial, owing to the following fact: Assuming both cyclicity of the trace and the anticommutation relations $\{ \gamma_*, \gamma^\mu \} = 0$, one can derive the identities
\begin{equation}
\left[ \prod_{k=0}^m (n-2k) \right] \tr \left( \gamma_* \gamma^{\mu_1} \cdots \gamma^{\mu_{2m}} \right) = 0 \eqend{,}
\end{equation}
which for $m = 2$ and $n \neq 4$ is inconsistent with the trace~\eqref{gamma_trace} $\tr \left( \gamma_* \gamma^{\mu\nu\rho\sigma} \right) = - 4 \mathi \epsilon^{\mu\nu\rho\sigma}$. To our knowledge, there are the following consistent prescriptions:
\begin{enumerate}
\item The original proposal by Breitenlohner and Maison~\cite{breitenlohnermaison1977}, following t'Hooft and Veltman~\cite{thooftveltman1972}: the matrix $\gamma_*$ is taken to be $- \frac{\mathi}{4!} \epsilon_{\mu\nu\lambda\rho} \gamma^\mu \gamma^\nu \gamma^\lambda \gamma^\rho$ also in $n$ dimensions, where $\epsilon_{\mu\nu\lambda\rho}$ remains four-dimensional (i.e., it vanishes when contracted with an object whose indices belong to the $(n-4)$-dimensional subspace). Consequently, $\gamma_*$ anticommutes with the first four $\gamma$ matrices but commutes with the other $n-4$ ones. This necessitates splitting Lorentz indices into four- and $(n-4)$-dimensional ones, but preserves the cyclicity of the trace. Because of the breaking of $n$-dimensional Lorentz invariance, further finite renormalisations may be necessary to preserve Ward identities.
\item One keeps $\gamma_*$ anticommuting with all $\gamma$ matrices, but drops the cyclic property of the trace~\cite{kreimer1990,koernerkreimerschilcher1992}. This can be realised by embedding the four-dimensional $\gamma$ matrix algebra in an infinite-dimensional one, and it can be shown that non-cyclicity is only relevant for traces containing an odd number of $\gamma_*$ and at least six $\gamma$ matrices. An advantage of this prescription is the preservation of $n$-dimensional Lorentz invariance.
\item The anticommutation relation $\{ \gamma_*, \gamma^\mu \} = 0$ is generalised to allow for a non-vanishing right-hand side in $n \neq 4$ dimensions~\cite{thompsonhu1985}. This preserves both the cyclicity of the trace and $n$-dimensional Lorentz invariance, but has the disadvantage of complicating the algebra by introducing new fully antisymmetric tensors, which only after renormalisation and the physical limit $n \to 4$ reduce to the $\epsilon$ tensor.
\end{enumerate}
Further proposals, such as dimensional reduction or abandoning associativity of the $\gamma$ matrix products, either have been shown to be inconsistent, or their consistency has not been proven, apart from one-loop calculations or in special cases~\cite{bonneau1980a,bonneau1980b,bonneau1981,baikovilyin1991}.

From their $\gamma_*$ commutation relations, it follows that Bonora et al.\ use the Breitenlohner--Maison prescription. However, we are not convinced that this prescription is used consistently in their work. Namely, it seems that they perform the $\gamma$ matrix algebra in $4$ dimensions, before regularising the integral and introducing the $(n-4)$-dimensional momentum components. There is a footnote in their first article~\cite{bonoraetal2014}, stating that doing otherwise would give a wrong result for the parity-even part of the anomaly; but as explained above the Breitenlohner--Maison prescription in general needs additional finite renormalisations to restore Ward identities which are broken because of the breaking of $n$-dimensional Lorentz invariance. As we have seen, also Hadamard subtraction needs to be supplemented by additional finite renormalisations to ensure a conserved renormalised stress tensor operator, and that this additional renormalisation not only removes the parity-odd term from the trace, but also changes the coefficients of the parity-even terms. Therefore, it seems to us that a calculation in dimensional regularisation, adhering strictly to one of the above consistent possibilities for the treatment of $\gamma_*$, possibly together with an additional finite renormalisation to restore Ward identities (i.e., conservation of the renormalised stress tensor), needs to be done and should give a result that coincides with ours (and others, such as the one by Bastianelli and Martelli~\cite{bastianellimartelli2016}). A new calculation by Bonora et el.~\cite{bonoraetal2018b} using ``axial gravity'' seems to confirm their result, independently of their previous calculations. However, the conservation of the stress tensor was not checked in~\cite{bonoraetal2018b}, and it is not clear to us whether the Wick rotation they perform in order to calculate the ``axial gravity'' heat kernel coefficients belongs to one of the two consistent formalisms presented above.

A further interesting extension to our results would be to consider also a non-vanishing background gauge field, as done by Bastianelli and Broccoli~\cite{bastianellibroccoli2019} using Pauli--Villars regularisation. The anomalous non-vanishing divergence of the fermion current has been treated using Hadamard subtraction in~\cite{zahn2014b}, and the extension to the trace anomaly is straightforward but lengthy.

\begin{acknowledgments}
We thank L.~Bonora and F.~Bastianelli for comments and references.
\end{acknowledgments}

\appendix

\section{Coincidence limits}
\label{appendix_coincidence}

In addition to the coincidence limits already given in equation~\eqref{spinor_parametrix_coincidence}, we also need
\begin{splitequation}
&\lim_{x' \to x} \nabla_{(\mu} \nabla_{\nu)} \mathcal{V}_0^{(0)}(x,x') \\
&= \frac{1}{4320} \bigg[ 72 \nabla_\mu \nabla_\nu R - 36 \nabla^2 R_{\mu\nu} + 30 R_{\mu\nu} R + 48 R_\mu^\rho R_{\nu\rho} - 24 R^{\rho\sigma} R_{\mu\rho\nu\sigma} + 13 R_{\mu}{}^{\rho\sigma\tau} R_{\nu\rho\sigma\tau} \bigg] \unitmatrix \\
&\quad+ \frac{1}{24} \left[ \nabla_\rho \nabla_{(\mu} R_{\nu)\sigma} + R_{(\mu}^\tau R_{\nu)\rho\sigma\tau} + R_\rho^\tau R_{\sigma(\mu\nu)\tau} \right] \gamma^{\rho\sigma} - \frac{1}{192} R_{\alpha\beta\rho(\mu} R^\rho{}_{\nu)\gamma\delta} \, \gamma^{\alpha\beta\gamma\delta} \eqend{,}
\end{splitequation}
\begin{splitequation}
&\lim_{x' \to x} \nabla_{(\mu} \nabla_\nu \nabla_{\rho)} \mathcal{V}_0^{(0)}(x,x') \\
&= \frac{1}{480} \nabla_{(\mu} \left[ 7 \nabla_\nu \nabla_{\rho)} R - 6 \nabla^2 R_{\nu\rho)} + 5 R R_{\nu\rho)} + 4 R^\alpha{}_{\nu\rho)}{}^\beta R_{\alpha\beta} + 8 R^\alpha_\nu R_{\rho)\alpha} \right] \unitmatrix \\
&\quad+ \frac{1}{960} \left[ 9 \nabla_{(\mu} \left( R_\nu{}^{\alpha\beta\gamma} R_{\rho)\alpha\beta\gamma} \right) - 8 R_{\gamma(\alpha\beta)(\mu} \nabla^\gamma R^\alpha{}_{\nu\rho)}{}^\beta \right] \unitmatrix - \frac{1}{64} R_{(\mu}{}^{\sigma\alpha\beta} \nabla_\nu R_{\rho)\sigma}{}^{\gamma\delta} \gamma^{\alpha\beta\gamma\delta} \\
&\quad+ \frac{1}{240} \bigg[ - 9 \nabla_{(\mu} \nabla_\nu \nabla^\beta R_{\rho)}^\alpha + 5 R_{(\mu\nu} \nabla^\alpha R_{\rho)}^\beta - R_{\gamma(\mu} \nabla_\nu R_{\rho)}{}^{\gamma\alpha\beta} \\
&\qquad\qquad+ 15 R^{\alpha\beta\gamma}{}_{(\mu} \nabla_\nu R_{\rho)\gamma} - 9 R^\alpha{}_{(\mu\nu}{}^\gamma \nabla_{\rho)} R^\beta_\gamma - 9 R^\beta{}_{(\mu\nu}{}^\gamma \nabla_{|\gamma|} R_{\rho)}^\alpha \\
&\qquad\qquad+ 3 R^{\alpha\beta}{}_{\gamma(\mu} \nabla^\gamma R_{\nu\rho)} + 2 R_{\gamma(\nu\rho}{}^\delta \nabla^\gamma R_{\mu)\delta}{}^{\alpha\beta} - 3 R^\alpha{}_{\delta\gamma(\mu} \nabla_\nu R_{\rho)}{}^{\gamma\delta\beta} \bigg] \gamma_{\alpha\beta} \eqend{,}
\end{splitequation}
and
\begin{splitequation}
&\lim_{x' \to x} \nabla_\mu \mathcal{V}_0^{(1)}(x,x') \\
&= - \frac{1}{11520} \nabla_\mu \left[ 7 R_{\alpha\beta\gamma\delta} R^{\alpha\beta\gamma\delta} + 8 R_{\alpha\beta} R^{\alpha\beta} - 5 R^2 + 12 \nabla^2 R \right] \unitmatrix + \frac{1}{1536} \nabla_\mu \left( R_{\alpha\beta\nu\rho} R_{\gamma\delta}{}^{\nu\rho} \right) \gamma^{\alpha\beta\gamma\delta} \\
&\quad+ \frac{1}{5760} \bigg[ - 12 \nabla^2 \nabla_\alpha R_{\beta\mu} + 10 R \nabla_\alpha R_{\mu\beta} - 3 R_{\mu\nu\alpha\beta} \nabla^\nu R + 4 R_{\mu\gamma} \nabla_\alpha R_\beta^\gamma + 4 R^{\nu\gamma} \nabla_\gamma R_{\mu\nu\alpha\beta} \\
&\qquad\quad+ 24 R_{\mu\nu\beta}{}^\gamma \nabla_{[\gamma} R_{\alpha]}^\nu - 12 R_{\alpha\beta\gamma\delta} \nabla^\gamma R_\mu^\delta + 6 R_{\alpha\gamma\delta\nu} \nabla_\mu R_\beta{}^{\gamma\delta\nu} + 4 R_{\mu\nu\gamma\delta} \nabla^\nu R_{\alpha\beta}{}^{\gamma\delta} \bigg] \gamma^{\alpha\beta} \eqend{.}
\end{splitequation}

\section{Two-component fermions}
\label{appendix_2comp}

For two-component fermions, we use the conventions of the review~\cite{dreinerhabermartin2010} except for the overall sign of the metric and (as a consequence) of $\bar{\sigma}^\mu$. For the sake of readability, we however do not explicitly show the spinor indices but stick with a matrix notation. We choose a representation of the $\gamma$ matrices of the block-diagonal form
\begin{equation}
\gamma^\mu = \begin{pmatrix} 0 & \sigma^\mu \\ \bar{\sigma}^\mu & 0 \end{pmatrix}
\end{equation}
with the curved-space $\sigma$ matrices $\sigma^\mu$ and $\bar{\sigma}^\mu$ obtained from the constant flat-space Pauli matrices $\vec{\sigma}_i$ using the frame field:
\begin{equations}[2comp_sigma_curved]
\sigma^\mu &\equiv g^{\mu\nu} e_\nu{}^a \eta_{ab} \left( \unitmatrix, \vec{\sigma} \right)^b = \epsilon \left( \bar{\sigma}^\mu \right)^\text{T} \epsilon \eqend{,} \\
\bar{\sigma}^\mu &\equiv g^{\mu\nu} e_\nu{}^a \eta_{ab} \left( - \unitmatrix, \vec{\sigma} \right)^b = \epsilon \left( \sigma^\mu \right)^\text{T} \epsilon \eqend{.}
\end{equations}
where $\epsilon = \left( \begin{smallmatrix} 0 & 1 \\ - 1 & 0 \end{smallmatrix} \right) = - \epsilon^\text{T}$ is the spin metric for two-component fermions with $\epsilon^2 = - \unitmatrix$. For the product of two of these matrices, one has
\begin{equation}
\label{2comp_sigma_product}
\bar{\sigma}^\mu \sigma^\nu = g^{\mu\nu} \unitmatrix - \frac{\mathi}{2} \epsilon^{\mu\nu\rho\sigma} \bar{\sigma}_\rho \sigma_\sigma \eqend{,} \qquad \sigma^\mu \bar{\sigma}^\nu = g^{\mu\nu} \unitmatrix + \frac{\mathi}{2} \epsilon^{\mu\nu\rho\sigma} \sigma_\rho \bar{\sigma}_\sigma \eqend{,}
\end{equation}
from which the analogue of the Clifford relations follow:
\begin{equation}
\label{2comp_sigma_clifford}
\bar{\sigma}^\mu \sigma^\nu + \bar{\sigma}^\nu \sigma^\mu = \sigma^\mu \bar{\sigma}^\nu + \sigma^\nu \bar{\sigma}^\mu = 2 g^{\mu\nu} \unitmatrix \eqend{.}
\end{equation}
One calculates easily that
\begin{equation}
\gamma_* = - \mathi \gamma^0 \gamma^1 \gamma^2 \gamma^3 = \begin{pmatrix} \unitmatrix & 0 \\ 0 & - \unitmatrix \end{pmatrix} \eqend{,}
\end{equation}
and it follows that for a left-handed Weyl fermion, a Dirac fermion satisfying $\psi = \mathcal{P}_+ \psi = \frac{1}{2} \left( \unitmatrix + \gamma_* \right) \psi$, one can isolate the upper two components $\chi$ and base the theory completely on $\chi$. Using that the Dirac adjoint $\bar{\psi}$ is given by $\bar{\psi} = \psi^\dagger \mathi \gamma^0$, the curved-space action~\eqref{weyl_action} becomes
\begin{equation}
\label{2comp_weyl_action}
S = - \int \chi^\dagger \overline{\mathcal{D}} \chi \sqrt{-g} \total^4 x
\end{equation}
where we define
\begin{equation}
\overline{\mathcal{D}} \equiv \mathi \bar{\sigma}^\mu \nabla_\mu \eqend{,} \qquad \mathcal{D} \equiv - \mathi \sigma^\mu \nabla_\mu \eqend{,} \qquad \overleftarrow{\overline{\mathcal{D}}} \equiv \mathi \overleftarrow{\nabla}_\mu \bar{\sigma}^\mu \eqend{,} \qquad \overleftarrow{\mathcal{D}} \equiv - \mathi \overleftarrow{\nabla}_\mu \sigma^\mu \eqend{,}
\end{equation}
and with the covariant derivative $\nabla_\mu$~\eqref{spinor_covd} acting on two-component spinors as
\begin{equation}
\label{2comp_spinor_covd}
\nabla_\mu \chi = \partial_\mu \chi + \frac{1}{4} \omega_{\mu\rho\sigma} \sigma^\rho \bar{\sigma}^\sigma \chi \eqend{,} \qquad \chi^\dagger \overleftarrow{\nabla}_\mu = \partial_\mu \chi^\dagger - \frac{1}{4} \omega_{\mu\rho\sigma} \chi^\dagger \bar{\sigma}^\rho \sigma^\sigma \eqend{.}
\end{equation}
In matrix notation, $\chi$ is treated as a column spinor while $\chi^\dagger$ is a row spinor. Since $\sigma^{\mu\dagger} = \sigma^\mu$ and $\bar{\sigma}^{\mu\dagger} = \bar{\sigma}^\mu$, the operators $\mathcal{D}$ and $\overline{\mathcal{D}}$ are formally self-adjoint, and from the action~\eqref{2comp_weyl_action} one sees that $\overline{\mathcal{D}}$ acts from the left on spinors, and from the right on cospinors. Using the relations~\eqref{2comp_sigma_curved} between the barred and unbarred $\sigma$ matrices and transposing to transform $\overline{\mathcal{D}}$ into $\mathcal{D}$, we find that $\mathcal{D}$ consequently acts from the right on the row spinor $( \epsilon \chi )^\text{T}$ and from the left on the column spinor $\epsilon \chi^{\dagger\text{T}} = - ( \chi^\dagger \epsilon )^\text{T}$. For them, the covariant derivatives~\eqref{2comp_spinor_covd} are given by
\begin{equation}
\label{2comp_spinor_covd2}
\nabla_\mu ( \epsilon \chi )^\text{T} = \partial_\mu ( \epsilon \chi )^\text{T} - \frac{1}{4} \omega_{\mu\rho\sigma} ( \epsilon \chi )^\text{T} \sigma^\rho \bar{\sigma}^\sigma \eqend{,} \qquad \nabla_\mu ( \epsilon \chi^{\dagger\text{T}} ) = \partial_\mu ( \epsilon \chi^{\dagger\text{T}} ) + \frac{1}{4} \omega_{\mu\rho\sigma} \bar{\sigma}^\rho \sigma^\sigma \epsilon \chi^{\dagger\text{T}} \eqend{,}
\end{equation}
where we have again used the relation~\eqref{2comp_sigma_curved} between the barred and unbarred $\sigma$ matrices.

Since the action remains Lorentz invariant, the stress tensor is obtained by varying the action~\eqref{2comp_weyl_action} with respect to the symmetric part of the frame field~\eqref{frame_var_sym_asym} as before, and we obtain
\begin{equation}
\label{2comp_weyl_stresstensor_def}
T^{\mu\nu} = \frac{\mathi}{2} \chi^\dagger \left[ \bar{\sigma}^{(\mu} \nabla^{\nu)} - \overleftarrow{\nabla}^{(\mu} \bar{\sigma}^{\nu)} \right] \chi + \frac{1}{2} g^{\mu\nu} \left[ \chi^\dagger \overleftarrow{\overline{\mathcal{D}}} \chi - \chi^\dagger \overline{\mathcal{D}} \chi \right] \eqend{.}
\end{equation}
The Feynman propagator, equal to the time-ordered two-point function in the state $\ket{W}$, is given by
\begin{equation}
G^\text{F}(x,x') \equiv - \mathi \bra{W} \mathcal{T} \chi(x) \chi^\dagger(x') \ket{W} \eqend{,}
\end{equation}
and fulfills 
\begin{equation}
\overline{\mathcal{D}} G^\text{F}(x,x') = \delta(x,x') \unitmatrix = - G^\text{F}(x,x') \overleftarrow{\overline{\mathcal{D}}}' \eqend{.}
\end{equation}
Again, we split it into a geometrically determined part $H^\text{F}$ and a smooth part $W$ which is state dependent, according to
\begin{equation}
G^\text{F}(x,x') = H^\text{F}(x,x') - \frac{\mathi}{8 \pi^2} W(x,x') \eqend{.}
\end{equation}
Analogous to the calculation with four-component spinors, we want to represent the para\-metrix $H$ as
\begin{equation}
\label{2comp_spinor_kg_parametrix}
H(x,x') = \mathcal{D} \mathcal{H}(x,x') \eqend{,}
\end{equation}
where $\mathcal{H}$ is the parametrix corresponding to the wave operator $P \equiv \overline{\mathcal{D}} \mathcal{D}$. Since $\mathcal{D}$ acts from the left on $\epsilon \chi^{\dagger\text{T}}$, we need to compute the corresponding commutator of covariant derivatives, and from the definitions~\eqref{2comp_spinor_covd} and~\eqref{2comp_spinor_covd2} we calculate
\begin{equation}
\label{2comp_covd_commutator}
\left[ \nabla_\mu, \nabla_\nu \right] \chi = \frac{1}{4} R_{\mu\nu\alpha\beta} \sigma^\alpha \bar{\sigma}^\beta \chi \eqend{,} \qquad \left[ \nabla_\mu, \nabla_\nu \right] ( \epsilon \chi^{\dagger\text{T}} ) = \frac{1}{4} R_{\mu\nu\alpha\beta} \bar{\sigma}^\alpha \sigma^\beta \epsilon \chi^{\dagger\text{T}} \eqend{.}
\end{equation}
Using the Clifford algebra relations~\eqref{2comp_sigma_clifford}, and the above commutator, the analogue of the calculation leading to~\eqref{spinor_hadamard_pmdef} then gives $P = \nabla^2 - \frac{1}{4} R$. As for Dirac fermions, this has the form of a spinorial Klein--Gordon operator, and the parametrix $\mathcal{H}$ is thus of the form
\begin{equation}
\label{2comp_hadamard_parametrix}
\mathcal{H}(x,x') \equiv - \frac{\mathi}{8 \pi^2} \left[ \frac{\mathcal{U}}{\sigma_\epsilon} + \mathcal{V} \ln\left( \mu^2 \sigma_\epsilon \right) \right] \eqend{,}
\end{equation}
with the same prescriptions as before (i.e., the Wightman prescription $\sigma_\epsilon = \sigma + \mathi \epsilon (t-t')$ for the two-point function and the Feynman prescription $\sigma_\epsilon = \sigma + \mathi \epsilon$ for the Feynman propagator). Completely analogous to the previous cases, the asymptotic expansions
\begin{equations}[2comp_hadamard_4d_spinorexpansion]
\mathcal{U} &= \mathcal{U}^{(0)} = \sqrt{\Delta} \, \mathcal{I} \eqend{,} \\
\mathcal{V} &= \sum_{k=0}^\infty \mathcal{V}^{(k)} \sigma^k
\end{equations}
with smooth biscalars $\mathcal{V}^{(k)}$ follow, where the two-component parallel propagator of cospinors $\mathcal{I}$ fulfills the analogue of the relation~\eqref{spinorparallel_relations}. With the same operator $Q_k$~\eqref{qk_def} as before, we again have $Q_0 \mathcal{U} = 0$ and obtain the recursion relations
\begin{equations}[2comp_hadamard_4d_spinor_recursion]
Q_{2k+4} \mathcal{V}^{(k+1)} &= - \frac{1}{k+1} P \mathcal{V}^{(k)} \eqend{,} \\
Q_{2k+4} \mathcal{W}^{(k+1)} &= - \frac{1}{k+1} \left[ P \mathcal{W}^{(k)} + Q_{4k+6} \mathcal{V}^{(k+1)} \right]
\end{equations}
subject to the boundary condition
\begin{equation}
\label{2comp_hadamard_4d_calv0_bdy}
Q_2 \mathcal{V}^{(0)} = - P \sqrt{\Delta} \, \mathcal{I} \eqend{,}
\end{equation}
and $\mathcal{V}$ is completely determined locally by these relations. As an auxiliary object we will below also use a second parametrix $\hat{\mathcal{H}}$ corresponding to the wave operator $\hat{P} \equiv \mathcal{D} \overline{\mathcal{D}} = \nabla^2 - \frac{1}{4} R$. The parametrix $\hat{\mathcal{H}}$ has the form analogous to~\eqref{2comp_hadamard_parametrix} with coefficients $\hat{\mathcal{U}}$ and $\hat{\mathcal{V}}$, which in turn admit the analogue of the expansion~\eqref{2comp_hadamard_4d_spinorexpansion} and the recursion relations~\eqref{2comp_hadamard_4d_spinor_recursion} and boundary condition~\eqref{2comp_hadamard_4d_calv0_bdy}, with $P$ replaced by $\hat{P}$.

Analogously to the case of four-component fermions, we write the stress tensor~\eqref{2comp_weyl_stresstensor_def} as
\begin{equation}
T^{\mu\nu} = \frac{1}{2} X^{\mu\nu} - \frac{1}{2} g^{\mu\nu} g_{\rho\sigma} X^{\rho\sigma}
\end{equation}
with
\begin{equation}
X^{\mu\nu} \equiv \mathi \chi^\dagger \left[ \bar{\sigma}^{(\mu} \nabla^{\nu)} - \overleftarrow{\nabla}^{(\mu} \bar{\sigma}^{\nu)} \right] \chi \eqend{.}
\end{equation}
The renormalised stress tensor operator $T^{\mu\nu}_\text{ren}$ is then defined as
\begin{equation}
T^{\mu\nu}_\text{ren} \equiv \frac{1}{2} X^{\mu\nu}_\text{ren} - \frac{1}{2} g^{\mu\nu} g_{\rho\sigma} X^{\rho\sigma}_\text{ren} \eqend{,}
\end{equation}
with the renormalised operator $X^{\mu\nu}_\text{ren}$ given by by point splitting and Hadamard subtraction,
\begin{equation}
\label{2comp_chimunu_pointsplit_def}
X^{\mu\nu}_\text{ren}(x) = - \mathi \tr \lim_{x' \to x} \left[ \bar{\sigma}^{(\mu} \nabla^{\nu)} \left( \chi(x) \chi^\dagger(x') - \mathi H(x,x') \right) - \left( \chi(x) \chi^\dagger(x') - \mathi H(x,x') \right) \overleftarrow{\nabla}^{(\mu'} \bar{\sigma}^{\nu')} \right] \eqend{,}
\end{equation}
where again $\tr$ denotes a trace in two-component spinor space and we have taken into account the minus sign from interchanging the two spinors. The renormalisation freedom is given by the same expression as before:
\begin{equation}
\label{2comp_ren_freedom}
X^{\mu\nu}_\text{ren} \to X^{\mu\nu}_\text{ren} + C^{\mu\nu} \unitmatrix \eqend{,} \qquad T^{\mu\nu}_\text{ren} \to T^{\mu\nu}_\text{ren} + \frac{1}{2} \left( C^{\mu\nu} - g^{\mu\nu} g_{\rho\sigma} C^{\rho\sigma} \right) \unitmatrix \eqend{,}
\end{equation}
where $C^{\mu\nu}$ is a symmetric tensor of dimension $4$ constructed from curvature tensors and their covariant derivatives.

The divergence and trace of the renormalised stress tensor operator are given by
\begin{equation}
\label{2comp_tmunu_ren_div_trace}
\nabla_\mu T^{\mu\nu}_\text{ren} = \frac{1}{2} \nabla_\mu X^{\mu\nu}_\text{ren} - \frac{1}{2} g_{\rho\sigma} \nabla^\nu X^{\rho\sigma}_\text{ren} \eqend{,} \qquad g_{\mu\nu} T^{\mu\nu}_\text{ren} = - \frac{3}{2} g_{\mu\nu} X^{\mu\nu}_\text{ren} \eqend{.}
\end{equation}
Using the point-split expression~\eqref{2comp_chimunu_pointsplit_def} we calculate
\begin{splitequation}
\label{2comp_xmunu_ren_div}
\nabla_\mu X^{\mu\nu}_\text{ren}(x) &= - \frac{1}{2} \tr \lim_{x' \to x} \Bigg[ \nabla^\nu \overline{\mathcal{D}} \left( \chi(x) \chi^\dagger(x') - \mathi H(x,x') \right) - \overline{\mathcal{D}} \left( \chi(x) \chi^\dagger(x') - \mathi H(x,x') \right) \overleftarrow{\nabla}^{\nu'} \\
&\qquad+ \mathi \bar{\sigma}^\nu \mathcal{D} \overline{\mathcal{D}} \left( \chi(x) \chi^\dagger(x') - \mathi H(x,x') \right) + \nabla^\nu \left( \chi(x) \chi^\dagger(x') - \mathi H(x,x') \right) \overleftarrow{\overline{\mathcal{D}}}' \\
&\qquad- \left( \chi(x) \chi^\dagger(x') - \mathi H(x,x') \right) \overleftarrow{\overline{\mathcal{D}}}' \overleftarrow{\nabla}^{\nu'} - \mathi \left( \chi(x) \chi^\dagger(x') - \mathi H(x,x') \right) \overleftarrow{\overline{\mathcal{D}}}' \overleftarrow{\mathcal{D}}' \bar{\sigma}^{\nu'} \Bigg] \\
&= \frac{\mathi}{2} \tr \lim_{x' \to x} \Bigg[ \nabla^\nu \overline{\mathcal{D}} H(x,x') - \overline{\mathcal{D}} H(x,x') \overleftarrow{\nabla}^{\nu'} + \mathi \bar{\sigma}^\nu \mathcal{D} \overline{\mathcal{D}} H(x,x') \\
&\qquad+ \nabla^\nu H(x,x') \overleftarrow{\overline{\mathcal{D}}}' - H(x,x') \overleftarrow{\overline{\mathcal{D}}}' \overleftarrow{\nabla}^{\nu'} - \mathi H(x,x') \overleftarrow{\overline{\mathcal{D}}}' \overleftarrow{\mathcal{D}}' \bar{\sigma}^{\nu'} \Bigg] \raisetag{2.4em}
\end{splitequation}
and
\begin{splitequation}
\label{2comp_xmunu_ren_trace}
g_{\mu\nu} X^{\mu\nu}_\text{ren} &= - \tr \lim_{x' \to x} \left[ \overline{\mathcal{D}} \left( \chi(x) \chi^\dagger(x') - \mathi H(x,x') \right) - \left( \chi(x) \chi^\dagger(x') - \mathi H(x,x') \right) \overleftarrow{\overline{\mathcal{D}}}' \right] \\
&= \mathi \tr \lim_{x' \to x} \left[ \overline{\mathcal{D}} H(x,x') - H(x,x') \overleftarrow{\overline{\mathcal{D}}}' \right] \eqend{,}
\end{splitequation}
where we used Synge's rule~\eqref{synge_rule} and the cyclicity of the trace and commuted covariant derivatives. Since the spinor field operator satisfies its equation of motion $\overline{\mathcal{D}} \chi = 0 = \chi^\dagger \overleftarrow{\mathcal{D}}$, only the terms containing the Hadamard parametrix remain. Completely analogous to the calculation for four-component spinors, we use the representation~\eqref{2comp_spinor_kg_parametrix} of the parametrix and the transport equations~\eqref{2comp_hadamard_4d_spinor_recursion},~\eqref{2comp_hadamard_4d_calv0_bdy} to calculate first
\begin{equation}
\overline{\mathcal{D}} H(x,x') = \overline{\mathcal{D}} \mathcal{D} \mathcal{H}(x,x') = - \frac{\mathi}{8 \pi^2} \sum_{k=0}^\infty Q_{4k+6} \mathcal{V}^{(k+1)} \sigma^k \eqend{,}
\end{equation}
and from this we obtain with the coincidence limits~\eqref{coincidence_geom_1} that
\begin{equations}
\lim_{x' \to x} \overline{\mathcal{D}} H(x,x') &= - \frac{3 \mathi}{4 \pi^2} \lim_{x' \to x} \mathcal{V}^{(1)} \eqend{,} \\
\lim_{x' \to x} \mathcal{D} \overline{\mathcal{D}} H(x,x') &= - \frac{1}{\pi^2} \sigma^\mu \lim_{x' \to x} \nabla_\mu \mathcal{V}^{(1)} \eqend{,} \\
\lim_{x' \to x} \nabla^\nu \overline{\mathcal{D}} H(x,x') &= - \frac{\mathi}{\pi^2} \lim_{x' \to x} \nabla^\nu \mathcal{V}^{(1)} \eqend{.}
\end{equations}

The remaining coincidence limits in~\eqref{2comp_xmunu_ren_div} and~\eqref{2comp_xmunu_ren_trace} all involve $H \overleftarrow{\overline{\mathcal{D}}}'$. To treat them, we add and subtract an expression involving the auxiliary parametrix $\hat{\mathcal{H}}$:
\begin{equation}
\label{2comp_add_subtract}
 H(x,x') \overleftarrow{\overline{\mathcal{D}}}' = \mathcal{D} \left( \mathcal{H}(x,x')  \overleftarrow{\overline{\mathcal{D}}}' + \overline{\mathcal{D}} \hat{\mathcal{H}}(x,x') \right) - \mathcal{D} \overline{\mathcal{D}} \hat{\mathcal{H}}(x,x') \eqend{.}
\end{equation}
We stress that both the original parametrix $\mathcal{H}$ and the auxiliary one $\hat{\mathcal{H}}$ are parametrices for left-handed two-component Weyl spinors only, and no right-handed spinors appear. The difference is that $\mathcal{H}$ is a parametrix for cospinors, while $\hat{\mathcal{H}}$ is a parametrix for spinors, as can be inferred from the operators $\mathcal{D}$ and $\overline{\mathcal{D}}$ acting on them, cf.\ the discussion at the beginning of this section. The two terms on the right-hand side of~\eqref{2comp_add_subtract} are smooth and can be separately computed. For the second term, we calculate, analogous to the above calculation for the parametrix $\mathcal{H}$, that
\begin{equation}
\label{2comp_ddhath}
\mathcal{D} \overline{\mathcal{D}} \hat{\mathcal{H}}(x,x') = - \frac{\mathi}{8 \pi^2} \sum_{k=0}^\infty Q_{4k+6} \hat{\mathcal{V}}^{(k+1)} \sigma^k \eqend{.}
\end{equation}
For the first term on the right-hand side of~\eqref{2comp_add_subtract}, we obtain by inserting the expansion~\eqref{2comp_hadamard_parametrix} that
\begin{splitequation}
\mathcal{H}(x,x') \overleftarrow{\overline{\mathcal{D}}}' + \overline{\mathcal{D}} \hat{\mathcal{H}}(x,x') &= - \frac{1}{8 \pi^2} \frac{1}{\sigma_\epsilon^2} \left( \mathcal{U} \bar{\sigma}^{\mu'} \nabla_{\mu'} \sigma + \bar{\sigma}^\mu \hat{\mathcal{U}} \nabla_\mu \sigma \right) \\
&\quad+ \frac{1}{8 \pi^2} \frac{1}{\sigma_\epsilon} \left[ \left( \mathcal{U} \overleftarrow{\nabla}_{\mu'} + \mathcal{V}^{(0)} \nabla_{\mu'} \sigma \right) \bar{\sigma}^{\mu'} + \bar{\sigma}^\mu \left( \nabla_\mu \hat{\mathcal{U}} + \hat{\mathcal{V}}^{(0)} \nabla_\mu \sigma \right) \right] \\
&\quad+ \frac{1}{8 \pi^2} \ln\left( \mu^2 \sigma_\epsilon \right) \sum_{k=0}^\infty \sigma^k \Bigg[ \left( \mathcal{V}^{(k)} \overleftarrow{\nabla}_{\mu'} + (k+1) \mathcal{V}^{(k+1)} \nabla_{\mu'} \sigma \right) \bar{\sigma}^{\mu'} \\
&\hspace{10em}+ \bar{\sigma}^\mu \left( \nabla_\mu \hat{\mathcal{V}}^{(k)} + (k+1) \hat{\mathcal{V}}^{(k+1)} \nabla_\mu \sigma \right) \Bigg] \\
&\quad+ \frac{1}{8 \pi^2} \sum_{k=0}^\infty \sigma^k \left[ \mathcal{V}^{(k+1)} \bar{\sigma}^{\mu'} \nabla_{\mu'} \sigma + \bar{\sigma}^\mu \hat{\mathcal{V}}^{(k+1)} \nabla_\mu \sigma \right] \eqend{.} \raisetag{2.4em}
\end{splitequation}
Using the analogue of~\eqref{gamma_xxs_parallel} for $\sigma$ matrices, $\bar{\sigma}^\nu g_\nu{}^{\beta'} \mathcal{I} = \mathcal{I} \bar{\sigma}^{\beta'}$, it follows that the most singular term proportional to $\sigma_\epsilon^{-2}$ vanishes. For the singular terms proportional to $\sigma_\epsilon^{-1}$ and $\ln\left( \mu^2 \sigma_\epsilon \right)$ one derives a transport equation with vanishing boundary term, and since the unique smooth solution to such an equation vanishes, these terms vanish as well. The calculation is lengthy but completely analogous to the calculation leading to~\eqref{eq:DH_0+H_0D_smooth} in the case of four-component spinors, such that we do not show any details. We are thus left with
\begin{equation}
\mathcal{H}(x,x') \overleftarrow{\overline{\mathcal{D}}}' + \overline{\mathcal{D}} \hat{\mathcal{H}}(x,x') = \frac{1}{8 \pi^2} \sum_{k=0}^\infty \sigma^k \left[ \mathcal{V}^{(k+1)} \bar{\sigma}^{\mu'} \nabla_{\mu'} \sigma + \bar{\sigma}^\mu \hat{\mathcal{V}}^{(k+1)} \nabla_\mu \sigma \right] \eqend{,}
\end{equation}
and inserting this and~\eqref{2comp_ddhath} into~\eqref{2comp_add_subtract} it follows that
\begin{splitequation}
H(x,x') \overleftarrow{\overline{\mathcal{D}}}' &= \frac{1}{8 \pi^2} \mathcal{D} \sum_{k=0}^\infty \sigma^k \left[ \mathcal{V}^{(k+1)} \bar{\sigma}^{\mu'} \nabla_{\mu'} \sigma + \bar{\sigma}^\mu \hat{\mathcal{V}}^{(k+1)} \nabla_\mu \sigma \right] + \frac{\mathi}{8 \pi^2} \sum_{k=0}^\infty Q_{4k+6} \hat{\mathcal{V}}^{(k+1)} \sigma^k \\
&= \frac{\mathi}{8 \pi^2} \sum_{k=0}^\infty \sigma^k \Bigg[ Q_{4k+6} \hat{\mathcal{V}}^{(k+1)} \\
&\hspace{4em}- \sigma^\rho \bar{\sigma}^\mu \left[ \nabla_\rho \hat{\mathcal{V}}^{(k+1)} \nabla_\mu \sigma + \hat{\mathcal{V}}^{(k+1)} \nabla_\mu \nabla_\rho \sigma + (k+1) \hat{\mathcal{V}}^{(k+2)} \nabla_\mu \sigma \nabla_\rho \sigma \right] \\
&\hspace{4em}- \sigma^\rho \left[ \nabla_\rho \mathcal{V}^{(k+1)} \nabla_{\mu'} \sigma + \mathcal{V}^{(k+1)} \nabla_{\mu'} \nabla_\rho \sigma + (k+1) \mathcal{V}^{(k+2)} \nabla_{\mu'} \sigma \nabla_\rho \sigma \right] \bar{\sigma}^{\mu'} \Bigg] \eqend{.}
\end{splitequation}
Using the coincidence limits~\eqref{coincidence_geom_1} and Synge's rule~\eqref{synge_rule}, we calculate
\begin{equations}
\lim_{x' \to x} H(x,x') \overleftarrow{\overline{\mathcal{D}}}' &= \frac{\mathi}{8 \pi^2} \lim_{x' \to x} \left[ 2 \hat{\mathcal{V}}^{(1)} + \sigma_\mu \mathcal{V}^{(1)} \bar{\sigma}^\mu \right] \eqend{,} \\
\begin{split}
\lim_{x' \to x} H(x,x') \overleftarrow{\overline{\mathcal{D}}}' \overleftarrow{\mathcal{D}}' &= - \frac{1}{8 \pi^2} \lim_{x' \to x} \left[ 2 \nabla_\nu \hat{\mathcal{V}}^{(1)} + \sigma_\nu \bar{\sigma}^\rho \nabla_\rho \hat{\mathcal{V}}^{(1)} + \sigma^\rho \nabla_\rho \mathcal{V}^{(1)} \bar{\sigma}_\nu + \sigma_\rho \nabla_\nu \mathcal{V}^{(1)} \bar{\sigma}^\rho \right] \sigma^\nu \\
&\quad+ \frac{1}{8 \pi^2} \nabla_\nu \lim_{x' \to x} \left[ 2 \hat{\mathcal{V}}^{(1)} + \sigma_\rho \mathcal{V}^{(1)} \bar{\sigma}^\rho \right] \sigma^\nu \eqend{,} \raisetag{1.8em}
\end{split} \\
\begin{split}
\lim_{x' \to x} H(x,x') \overleftarrow{\overline{\mathcal{D}}}' \overleftarrow{\nabla}^{\nu'} &= - \frac{\mathi}{8 \pi^2} \lim_{x' \to x} \left[ 2 \nabla^\nu \hat{\mathcal{V}}^{(1)} + \sigma^\nu \bar{\sigma}^\rho \nabla_\rho \hat{\mathcal{V}}^{(1)} + \sigma^\rho \nabla_\rho \mathcal{V}^{(1)} \bar{\sigma}^\nu + \sigma_\rho \nabla^\nu \mathcal{V}^{(1)} \bar{\sigma}^\rho \right] \\
&\quad+ \frac{\mathi}{8 \pi^2} \nabla^\nu \lim_{x' \to x} \left[ 2 \hat{\mathcal{V}}^{(1)} + \sigma_\rho \mathcal{V}^{(1)} \bar{\sigma}^\rho \right] \eqend{.} \raisetag{1.8em}
\end{split}
\end{equations}
Inserting these coincidence limits in the point-split expressions~\eqref{2comp_xmunu_ren_div} and~\eqref{2comp_xmunu_ren_trace} and using Synge's rule~\eqref{synge_rule}, for the divergence and trace of the renormalised stress tensor operator~\eqref{2comp_tmunu_ren_div_trace} we thus obtain
\begin{equations}[2comp_xtmunu_div_trace]
\begin{split}
\nabla_\mu T^{\mu\nu}_\text{ren} &= \frac{1}{16 \pi^2} \tr \lim_{x' \to x} \left[ - 4 \nabla^\nu \hat{\mathcal{V}}^{(1)} + \sigma^\nu \bar{\sigma}^\mu \nabla_\mu \hat{\mathcal{V}}^{(1)} + 2 \nabla^\nu \mathcal{V}^{(1)} + \bar{\sigma}^\nu \sigma^\mu \nabla_\mu \mathcal{V}^{(1)} \right] \\
&\quad+ \frac{1}{16 \pi^2} \tr \nabla_\mu \lim_{x' \to x} \left[ - g^{\mu\nu} \hat{\mathcal{V}}^{(1)} + \sigma^\mu \bar{\sigma}^\nu \hat{\mathcal{V}}^{(1)} - 9 g^{\mu\nu} \mathcal{V}^{(1)} \right] \eqend{,}
\end{split} \\
g_{\mu\nu} T^{\mu\nu}_\text{ren} &= - \frac{3}{8 \pi^2} \tr \lim_{x' \to x} \left[ 5 \mathcal{V}^{(1)} + \hat{\mathcal{V}}^{(1)} \right] \eqend{.}
\end{equations}
Again, only the second coefficient $\mathcal{V}^{(1)}$ and $\hat{\mathcal{V}}^{(1)}$ of each parametrix and their derivatives contribute. Taking the coincidence limits of the recursion~\eqref{2comp_hadamard_4d_spinor_recursion} and its derivative, we obtain as before (with the analogous relations for $\hat{\mathcal{V}}^{(1)}$)
\begin{equation}
\lim_{x' \to x} \mathcal{V}^{(1)} = - \frac{1}{4} \lim_{x' \to x} P \mathcal{V}^{(0)} \eqend{,} \qquad \lim_{x' \to x} \nabla_\mu \mathcal{V}^{(1)} = - \frac{1}{6} \lim_{x' \to x}  \nabla_\mu P \mathcal{V}^{(0)} \eqend{,}
\end{equation}
and thus need to determine the coincidence limits of $\mathcal{V}^{(0)}$, $\hat{\mathcal{V}}^{(0)}$ and their derivatives up to third order. This is again obtained by taking the coincidence limit of the boundary condition~\eqref{2comp_hadamard_4d_calv0_bdy} and its derivatives, using the coincidence limits~\eqref{coincidence_highorder}. While the spinor parallel propagator is now different, it still satisfies the same defining equation~\eqref{spinorparallel_relations} (with a two-by-two identity matrix), and coincidence limits of symmetrised derivatives only use this defining equation. However, for the non-symmetrised derivatives we need to use the commutation relation~\eqref{2comp_covd_commutator} instead of~\eqref{spinor_covd_commutator}, which changes the coincidence limits of $\mathcal{V}^{(0)}$, $\hat{\mathcal{V}}^{(0)}$ and their derivatives.

A calculation with xAct~\cite{xact}, using the Bianchi identities for the Riemann tensor gives
\begin{splitequation}
\lim_{x' \to x} \mathcal{V}^{(1)} &= \frac{1}{17280} \left( 32 R^{\alpha\beta\gamma\delta} R_{\alpha\beta\gamma\delta} - 24 R^{\alpha\beta} R_{\alpha\beta} + 15 R^2 - 36 \nabla^2 R \right) \unitmatrix \\
&\quad+ \frac{1}{768} R_{\alpha\beta\mu\nu} R_{\gamma\delta}{}^{\mu\nu} \bar{\sigma}^\alpha \sigma^\beta \bar{\sigma}^\gamma \sigma^\delta
\end{splitequation}
and
\begin{splitequation}
\lim_{x' \to x} \nabla_\mu \hat{\mathcal{V}}^{(1)} &= \frac{1}{11520} \nabla_\mu \left( 8 R^{\alpha\beta\gamma\delta} R_{\alpha\beta\gamma\delta} - 8 R^{\alpha\beta} R_{\alpha\beta} + 5 R^2 - 12 \nabla^2 R \right) \unitmatrix \\
&\quad+ \frac{1}{2880} \left[ 2 R^{\gamma\delta} \nabla_\gamma R_{\mu\delta\alpha\beta} + 2 R_\mu^\nu \nabla_{[\alpha} R_{\beta]\nu} + 5 R \nabla_{[\alpha} R_{\beta]\mu} - 6 \nabla^2 \nabla_{[\alpha} R_{\beta]\mu} \right] \bar{\sigma}^\alpha \sigma^\beta \\
&\quad+ \frac{1}{1440} \left[ 2 R_{\mu\gamma\delta[\alpha} \left( \nabla_{\beta]} R^{\gamma\delta} - \nabla^\delta R_{\beta]}^\gamma \right) - 3 R_{\alpha\beta\gamma\delta} \nabla^\gamma R_\mu^\delta \right] \bar{\sigma}^\alpha \sigma^\beta \\
&\quad+ \frac{1}{1440} \left[ 3 R_{\nu\delta\gamma[\alpha} \nabla^\nu R_{\beta]}{}^{\gamma\delta}{}_\mu - 2 R_{\mu}{}^{\gamma\delta\nu} \nabla_\nu R_{\gamma\delta\alpha\beta} \right] \bar{\sigma}^\alpha \sigma^\beta - \frac{1}{1920} R_{\mu\nu\alpha\beta} \nabla^\nu R \bar{\sigma}^\alpha \sigma^\beta \\
&\quad- \frac{1}{4608} \left[ 4 R_{\mu\nu\alpha\beta} \nabla_{[\gamma} R_{\delta]}^\nu - 4 R_{\mu\nu\gamma\delta} \nabla_{[\alpha} R_{\beta]}^\nu - 3 \nabla_\mu \left( R_{\alpha\beta\rho\sigma} R_{\gamma\delta}{}^{\rho\sigma} \right) \right] \bar{\sigma}^\alpha \sigma^\beta \bar{\sigma}^\gamma \sigma^\delta \eqend{,}
\end{splitequation}
and the corresponding expressions for $\hat{\mathcal{V}}^{(1)}$ are obtained by interchanging $\sigma$ and $\bar{\sigma}$. Using the product of $\sigma$ matrices~\eqref{2comp_sigma_product}, the cyclicity of the trace and $\tr \unitmatrix = 2$, we derive the following trace relations for the curved-space $\sigma$ matrices:
\begin{equations}
\tr\left( \sigma^\alpha \bar{\sigma}^\beta \right) &= 2 g^{\alpha\beta} \eqend{,} \\
\tr\left( \sigma^\alpha \bar{\sigma}^\beta \sigma^\gamma \bar{\sigma}^\delta \right) &= 2 \left( g^{\alpha\delta} g^{\beta\gamma} - g^{\alpha\gamma} g^{\beta\delta} + g^{\alpha\beta} g^{\gamma\delta} - \mathi \epsilon^{\alpha\beta\gamma\delta} \right) \eqend{,} \\
\begin{split}
\tr \left( \sigma^\alpha \bar{\sigma}^\beta \sigma^\gamma \bar{\sigma}^\delta \sigma^\mu \bar{\sigma}^\nu \right) &= - 2 g^{\alpha\beta} \left( 2 g^{\gamma[\mu} g^{\nu]\delta} - g^{\gamma\delta} g^{\mu\nu} \right) + 2 g^{\alpha\gamma} \left( 2 g^{\beta[\mu} g^{\nu]\delta} - g^{\beta\delta} g^{\mu\nu} \right) \\
&\quad- 2 g^{\alpha\delta} \left( 2 g^{\beta[\mu} g^{\nu]\gamma} - g^{\beta\gamma} g^{\mu\nu} \right) - 2 g^{\alpha\mu} \left( 2 g^{\beta[\gamma} g^{\delta]\nu} + g^{\beta\nu} g^{\gamma\delta} \right) \\
&\quad+ 2 g^{\alpha\nu} \left( 2 g^{\beta[\gamma} g^{\delta]\mu} + g^{\beta\mu} g^{\gamma\delta} \right) \\
&\quad- 2 \mathi \left( \epsilon^{\alpha\beta\gamma\delta} g^{\mu\nu} - 2 \epsilon^{\alpha\beta\gamma[\mu} g^{\nu]\delta} - 2 \epsilon^{\mu\nu\delta[\alpha} g^{\beta]\gamma} + \epsilon^{\mu\nu\gamma\delta} g^{\alpha\beta} \right) \eqend{.}
\end{split}
\end{equations}
Inserting the above coincidence limits in the expressions for the divergence and trace of the renormalised stress tensor operator~\eqref{2comp_xtmunu_div_trace} and taking the trace, we arrive at [using the Bianchi identities and the Weyl tensor identities~\eqref{weyl_4d_id} and~\eqref{weyl_4d_id2}]
\begin{equations}
\begin{split}
\nabla_\mu T^{\mu\nu}_\text{ren} &= \frac{1}{23040 \pi^2} \nabla^\nu \Bigg[ 23 R^{\alpha\beta\gamma\delta} R_{\alpha\beta\gamma\delta} + 40 R^{\alpha\beta} R_{\alpha\beta} - 25 R^2 + 60 \nabla^2 R \\
&\hspace{8em}- 45 \mathi R_{\alpha\beta\gamma\delta} (\star R)^{\alpha\beta\gamma\delta} \Bigg] \eqend{,}
\end{split} \\
g_{\mu\nu} T^{\mu\nu}_\text{ren} &= \frac{1}{3840 \pi^2} \left[ 13 R^{\alpha\beta\gamma\delta} R_{\alpha\beta\gamma\delta} + 24 R^{\alpha\beta} R_{\alpha\beta} - 15 R^2 + 36 \nabla^2 R - 30 \mathi R_{\alpha\beta\gamma\delta} (\star R)^{\alpha\beta\gamma\delta} \right] \eqend{.}
\end{equations}
These expressions exactly coincide with the ones obtained using the four-component formalism~\eqref{weyl_stresstensor_expectationdivtrace}, showing that (as expected) the description of chiral fermions using four- or two-component spinors are equivalent. It follows that the renormalisation freedom of the renormalised stress tensor~\eqref{tmunuren_freedom} with the same redefinition~\eqref{cmunu_redef} makes it covariantly conserved, and gives the result~\eqref{tmunu_traceanom_final} for its trace.

\addcontentsline{toc}{section}{References}
\bibliography{literature}

\providecommand{\href}[2]{#2}\begingroup\raggedright\begin{thebibliography}{10}

\bibitem{bonoraetal2014}
L.~Bonora, S.~Giaccari and B.~Lima~de Souza, \emph{{Trace anomalies in chiral
  theories revisited}},
  \href{http://dx.doi.org/10.1007/JHEP07(2014)117}{\emph{JHEP} {\bf 07} (2014)
  117}, [\href{http://arxiv.org/abs/1403.2606}{{\tt 1403.2606}}].

\bibitem{nakayama2012}
Y.~Nakayama, \emph{{CP-violating CFT and trace anomaly}},
  \href{http://dx.doi.org/10.1016/j.nuclphysb.2012.02.006}{\emph{Nucl.~Phys.~B}
  {\bf 859} (2012) 288}, [\href{http://arxiv.org/abs/1201.3428}{{\tt
  1201.3428}}].

\bibitem{nakayama2018}
Y.~Nakayama, \emph{{Realization of impossible anomalies}},
  \href{http://dx.doi.org/10.1103/PhysRevD.98.085002}{\emph{Phys.~Rev.~D} {\bf
  D98} (2018) 085002}, [\href{http://arxiv.org/abs/1804.02940}{{\tt
  1804.02940}}].

\bibitem{bastianellimartelli2016}
F.~Bastianelli and R.~Martelli, \emph{{On the trace anomaly of a Weyl
  fermion}}, \href{http://dx.doi.org/10.1007/JHEP11(2016)178}{\emph{JHEP} {\bf
  11} (2016) 178}, [\href{http://arxiv.org/abs/1610.02304}{{\tt 1610.02304}}].

\bibitem{bonoraetal2017}
L.~Bonora, M.~Cvitan, P.~Dominis~Prester, A.~Duarte~Pereira, S.~Giaccari and
  T.~{\v S}temberga, \emph{{Axial gravity, massless fermions and trace
  anomalies}},
  \href{http://dx.doi.org/10.1140/epjc/s10052-017-5071-7}{\emph{Eur.~Phys.~J.~C}
  {\bf 77} (2017) 511}, [\href{http://arxiv.org/abs/1703.10473}{{\tt
  1703.10473}}].

\bibitem{bonoraetal2018}
L.~Bonora, M.~Cvitan, P.~D. Prester, A.~D. Pereira, S.~Giaccari and T.~{\v
  S}temberga, \emph{{Pontryagin trace anomaly}},
  \href{http://dx.doi.org/10.1051/epjconf/201818202100}{\emph{EPJ~Web~Conf.}
  {\bf 182} (2018) 02100}.

\bibitem{zahn2014b}
J.~Zahn, \emph{{Locally covariant chiral fermions and anomalies}},
  \href{http://dx.doi.org/10.1016/j.nuclphysb.2014.11.008}{\emph{Nucl.~Phys.~B}
  {\bf 890} (2015) 1}, [\href{http://arxiv.org/abs/1407.1994}{{\tt
  1407.1994}}].

\bibitem{hollandswald2015}
S.~Hollands and R.~M. Wald, \emph{{Quantum fields in curved spacetime}},
  \href{http://dx.doi.org/10.1016/j.physrep.2015.02.001}{\emph{Phys.~Rept.}
  {\bf 574} (2015) 1}, [\href{http://arxiv.org/abs/1401.2026}{{\tt
  1401.2026}}].

\bibitem{mtw1973}
C.~Misner, K.~Thorne and J.~A. Wheeler, \emph{{Gravitation}}.
\newblock
  \href{http://www.worldcat.org/search?q=isbn:9780716703440}{W.~H.~Freeman, San
  Francisco, 1973}.

\bibitem{hadamard1932}
J.~Hadamard, \emph{{Le probl{\`e}me de Cauchy et les {\'e}quations aux
  d{\'e}riv{\'e}es partielles lin{\'e}aires hyperboliques}}.
\newblock
  \href{http://www.worldcat.org/search?q=isbn:978-2-87647-300-3}{{Hermann et
  Cie.}, Paris, France, 1932}.
\newblock (in French).

\bibitem{baerginouxpfaeffle2007}
C.~B{\"a}r, N.~Ginoux and F.~Pf{\"a}ffle, \emph{{Wave Equations on Lorentzian
  Manifolds and Quantization}}.
\newblock {European Mathematical Society Publishing House}, Z{\"u}rich,
  Switzerland, 2007.

\bibitem{fullingnarcowichwald1981}
S.~A. Fulling, F.~J. Narcowich and R.~M. Wald, \emph{{Singularity Structure of
  the Two-Point Function in Quantum Field Theory in Curved Spacetime, II}},
  \href{http://dx.doi.org/10.1016/0003-4916(81)90098-1}{\emph{Annals~Phys.}
  {\bf 136} (1981) 243}.

\bibitem{fullingsweenywald1978}
S.~A. Fulling, M.~Sweeny and R.~M. Wald, \emph{{Singularity structure of the
  two-point function in quantum field theory in curved spacetime}},
  \href{http://dx.doi.org/10.1007/BF01196934}{\emph{Commun.~Math.~Phys.} {\bf
  63} (1978) 257}.

\bibitem{sahlmannverch2001}
H.~Sahlmann and R.~Verch, \emph{{Microlocal spectrum condition and Hadamard
  form for vector-valued quantum fields in curved spacetime}},
  \href{http://dx.doi.org/10.1142/S0129055X01001010}{\emph{Rev.~Math.~Phys.}
  {\bf 13} (2001) 1203}, [\href{http://arxiv.org/abs/math-ph/0008029}{{\tt
  math-ph/0008029}}].

\bibitem{sanders2010}
K.~Sanders, \emph{{The locally covariant Dirac field}},
  \href{http://dx.doi.org/10.1142/S0129055X10003990}{\emph{Rev.~Math.~Phys.}
  {\bf 22} (2010) 381}, [\href{http://arxiv.org/abs/0911.1304}{{\tt
  0911.1304}}].

\bibitem{sahlmannverch2000}
H.~Sahlmann and R.~Verch, \emph{{Passivity and microlocal spectrum condition}},
  \href{http://dx.doi.org/10.1007/s002200000297}{\emph{Commun.~Math.~Phys.}
  {\bf 214} (2000) 705}, [\href{http://arxiv.org/abs/math-ph/0002021}{{\tt
  math-ph/0002021}}].

\bibitem{parkertoms2009}
L.~Parker and D.~Toms, \emph{{Quantum Field Theory in Curved Spacetime:
  Quantized Fields and Gravity}}.
\newblock
  \href{http://www.worldcat.org/search?q=isbn:978-0521877879}{{Cambridge
  University Press}, Cambridge, UK, 2009}.

\bibitem{pinamonti2009}
N.~Pinamonti, \emph{{Conformal Generally Covariant Quantum Field Theory: The
  Scalar Field and its Wick Products}},
  \href{http://dx.doi.org/10.1007/s00220-009-0780-x}{\emph{Commun.~Math.~Phys.}
  {\bf 288} (2009) 1117}, [\href{http://arxiv.org/abs/0806.0803}{{\tt
  0806.0803}}].

\bibitem{olbermann2007}
H.~Olbermann, \emph{{States of low energy on Robertson--Walker spacetimes}},
  \href{http://dx.doi.org/10.1088/0264-9381/24/20/007}{\emph{Class.~Quant.~Grav.}
  {\bf 24} (2007) 5011}, [\href{http://arxiv.org/abs/0704.2986}{{\tt
  0704.2986}}].

\bibitem{thembrum2013}
K.~Them and M.~Brum, \emph{{States of low energy in homogeneous and
  inhomogeneous expanding spacetimes}},
  \href{http://dx.doi.org/10.1088/0264-9381/30/23/235035}{\emph{Class.~Quant.~Grav.}
  {\bf 30} (2013) 235035}, [\href{http://arxiv.org/abs/1302.3174}{{\tt
  1302.3174}}].

\bibitem{pirk1993}
K.-T. Pirk, \emph{{Hadamard states and adiabatic vacua}},
  \href{http://dx.doi.org/10.1103/PhysRevD.48.3779}{\emph{Phys.~Rev.~D} {\bf
  48} (1993) 3779}, [\href{http://arxiv.org/abs/gr-qc/9211003}{{\tt
  gr-qc/9211003}}].

\bibitem{hollands2001}
S.~Hollands, \emph{{The Hadamard Condition for Dirac Fields and Adiabatic
  States on Robertson--Walker Spacetimes}},
  \href{http://dx.doi.org/10.1007/s002200000350}{\emph{Commun.~Math.~Phys.}
  {\bf 216} (2001) 635}, [\href{http://arxiv.org/abs/gr-qc/9906076}{{\tt
  gr-qc/9906076}}].

\bibitem{junkerschrohe2002}
W.~Junker and E.~Schrohe, \emph{{Adiabatic Vacuum States on General Spacetime
  Manifolds: Definition, Construction, and Physical Properties}},
  \href{http://dx.doi.org/10.1007/s000230200001}{\emph{Ann.~H.~Poincar{\'e}}
  {\bf 3} (2002) 1113}, [\href{http://arxiv.org/abs/math-ph/0109010}{{\tt
  math-ph/0109010}}].

\bibitem{brunettifredenhagen2000}
R.~Brunetti and K.~Fredenhagen, \emph{{Microlocal Analysis and Interacting
  Quantum Field Theories: Renormalization on Physical Backgrounds}},
  \href{http://dx.doi.org/10.1007/s002200050004}{\emph{Commun.~Math.~Phys.}
  {\bf 208} (2000) 623}, [\href{http://arxiv.org/abs/math-ph/9903028}{{\tt
  math-ph/9903028}}].

\bibitem{fewsterverch2013}
C.~J. Fewster and R.~Verch, \emph{{The necessity of the Hadamard condition}},
  \href{http://dx.doi.org/10.1088/0264-9381/30/23/235027}{\emph{Class.~Quant.~Grav.}
  {\bf 30} (2013) 235027}, [\href{http://arxiv.org/abs/1307.5242}{{\tt
  1307.5242}}].

\bibitem{synge1960}
J.~L. Synge, \emph{{Relativity: The General Theory}}.
\newblock
  \href{http://www.worldcat.org/search?q=isbn:978-0720400663}{{North-Holland},
  Amsterdam, The Netherlands, 1960}.

\bibitem{moretti2000}
V.~Moretti, \emph{{Proof of the Symmetry of the Off-Diagonal
  Hadamard/Seeley--deWitt's Coefficients in {$C^\infty$} Lorentzian Manifolds
  by a ``Local Wick Rotation''}},
  \href{http://dx.doi.org/10.1007/s002200000202}{\emph{Commun.~Math.~Phys.}
  {\bf 212} (2000) 165}, [\href{http://arxiv.org/abs/gr-qc/9908068}{{\tt
  gr-qc/9908068}}].

\bibitem{hollandswald2005}
S.~Hollands and R.~M. Wald, \emph{{Conservation of the stress tensor in
  interacting quantum field theory in curved spacetimes}},
  \href{http://dx.doi.org/10.1142/S0129055X05002340}{\emph{Rev.~Math.~Phys.}
  {\bf 17} (2005) 227}, [\href{http://arxiv.org/abs/gr-qc/0404074}{{\tt
  gr-qc/0404074}}].

\bibitem{decaninifolacci2008}
Y.~D{\'e}canini and A.~Folacci, \emph{{Hadamard renormalization of the
  stress-energy tensor for a quantized scalar field in a general spacetime of
  arbitrary dimension}},
  \href{http://dx.doi.org/10.1103/PhysRevD.78.044025}{\emph{Phys.~Rev.~D} {\bf
  78} (2008) 044025}, [\href{http://arxiv.org/abs/gr-qc/0512118}{{\tt
  gr-qc/0512118}}].

\bibitem{dewittbrehme1960}
B.~S. DeWitt and R.~W. Brehme, \emph{{Radiation damping in a gravitational
  field}},
  \href{http://dx.doi.org/10.1016/0003-4916(60)90030-0}{\emph{Ann.~Phys.} {\bf
  9} (1960) 220}.

\bibitem{dirac1934}
P.~A.~M. Dirac, \emph{{Discussion of the infinite distribution of electrons in
  the theory of the positron}},
  \href{http://dx.doi.org/10.1017/S030500410001656X}{\emph{Proc.~Camb.~Phil.~Soc.}
  {\bf 30} (1934) 150}.

\bibitem{schwinger1951}
J.~S. Schwinger, \emph{{The Theory of Quantized Fields. I}},
  \href{http://dx.doi.org/10.1103/PhysRev.82.914}{\emph{Phys.~Rev.} {\bf 82}
  (1951) 914}. [,132(1951)].

\bibitem{dewitt1975}
B.~S. DeWitt, \emph{{Quantum field theory in curved spacetime}},
  \href{http://dx.doi.org/10.1016/0370-1573(75)90051-4}{\emph{Phys.~Rept.} {\bf
  19} (1975) 295}.

\bibitem{christensen1976}
S.~M. Christensen, \emph{{Vacuum expectation value of the stress tensor in an
  arbitrary curved background: The covariant point-separation method}},
  \href{http://dx.doi.org/10.1103/PhysRevD.14.2490}{\emph{Phys.~Rev.~D} {\bf
  14} (1976) 2490}.

\bibitem{christensen1978}
S.~M. Christensen, \emph{{Regularization, renormalization, and covariant
  geodesic point separation}},
  \href{http://dx.doi.org/10.1103/PhysRevD.17.946}{\emph{Phys.~Rev.~D} {\bf 17}
  (1978) 946}.

\bibitem{uehling1935}
E.~A. Uehling, \emph{{Polarization effects in the positron theory}},
  \href{http://dx.doi.org/10.1103/PhysRev.48.55}{\emph{Phys.~Rev.} {\bf 48}
  (1935) 55}.

\bibitem{fewsterverch2012}
C.~J. Fewster and R.~Verch, \emph{{Dynamical Locality and Covariance: What
  Makes a Physical Theory the Same in all Spacetimes?}},
  \href{http://dx.doi.org/10.1007/s00023-012-0165-0}{\emph{Ann.~H.~Poincar{\'e}}
  {\bf 13} (2012) 1613}, [\href{http://arxiv.org/abs/1106.4785}{{\tt
  1106.4785}}].

\bibitem{fewster2018}
C.~J. Fewster, \emph{{The art of the state}},
  \href{http://dx.doi.org/10.1142/S0218271818430071}{\emph{Int.~J.~Mod.~Phys.~D}
  {\bf 27} (2018) 1843007}, [\href{http://arxiv.org/abs/1803.06836}{{\tt
  1803.06836}}].

\bibitem{hollandswald2001}
S.~Hollands and R.~M. Wald, \emph{{Local Wick Polynomials and Time Ordered
  Products of Quantum Fields in Curved Spacetime}},
  \href{http://dx.doi.org/10.1007/s002200100540}{\emph{Commun.~Math.~Phys.}
  {\bf 223} (2001) 289}, [\href{http://arxiv.org/abs/gr-qc/0103074}{{\tt
  gr-qc/0103074}}].

\bibitem{hollandswald2002}
S.~Hollands and R.~M. Wald, \emph{{Existence of Local Covariant Time Ordered
  Products of Quantum Fields in Curved Spacetime}},
  \href{http://dx.doi.org/10.1007/s00220-002-0719-y}{\emph{Commun.~Math.~Phys.}
  {\bf 231} (2002) 309}, [\href{http://arxiv.org/abs/gr-qc/0111108}{{\tt
  gr-qc/0111108}}].

\bibitem{poissonpoundvega2011}
E.~Poisson, A.~Pound and I.~Vega, \emph{{The Motion of point particles in
  curved spacetime}},
  \href{http://dx.doi.org/10.12942/lrr-2011-7}{\emph{Living~Rev.~Rel.} {\bf 14}
  (2011) 7}, [\href{http://arxiv.org/abs/1102.0529}{{\tt 1102.0529}}].

\bibitem{moretti2003}
V.~Moretti, \emph{{Comments on the Stress-Energy Tensor Operator in Curved
  Spacetime}},
  \href{http://dx.doi.org/10.1007/s00220-002-0702-7}{\emph{Commun.~Math.~Phys.}
  {\bf 232} (2003) 189}, [\href{http://arxiv.org/abs/gr-qc/0109048}{{\tt
  gr-qc/0109048}}].

\bibitem{decaninifolacci2006}
Y.~D{\'e}canini and A.~Folacci, \emph{{Off-diagonal coefficients of the
  DeWitt-Schwinger and Hadamard representations of the Feynman propagator}},
  \href{http://dx.doi.org/10.1103/PhysRevD.73.044027}{\emph{Phys.~Rev.~D} {\bf
  73} (2006) 044027}, [\href{http://arxiv.org/abs/gr-qc/0511115}{{\tt
  gr-qc/0511115}}].

\bibitem{forgerroemer2004}
M.~Forger and H.~R{\"o}mer, \emph{{Currents and the energy-momentum tensor in
  classical field theory: a fresh look at an old problem}},
  \href{http://dx.doi.org/10.1016/j.aop.2003.08.011}{\emph{Annals~Phys.} {\bf
  309} (2004) 306}, [\href{http://arxiv.org/abs/hep-th/0307199}{{\tt
  hep-th/0307199}}].

\bibitem{kennedy1981}
A.~D. Kennedy, \emph{{Clifford algebras in $2\omega$ dimensions}},
  \href{http://dx.doi.org/10.1063/1.525069}{\emph{J.~Math.~Phys.} {\bf 22}
  (1981) 1330}.

\bibitem{zahn2014a}
J.~Zahn, \emph{{The renormalized locally covariant Dirac field}},
  \href{http://dx.doi.org/10.1142/S0129055X13300124}{\emph{Rev.~Math.~Phys.}
  {\bf 26} (2014) 1330012}, [\href{http://arxiv.org/abs/1210.4031}{{\tt
  1210.4031}}].

\bibitem{xact}
J.~M. Mart{\'\i}n-Garc{\'\i}a, A.~Garc{\'\i}a-Parrado, A.~Stecchina,
  B.~Wardell, C.~Pitrou, D.~Brizuela et~al., ``{xAct: Efficient tensor computer
  algebra for the Wolfram Language }.''
  \href{http://www.xact.es}{http://www.xact.es}, 2018.

\bibitem{lovelock1970}
D.~Lovelock, \emph{{Dimensionally dependent identities}},
  \href{http://dx.doi.org/10.1017/S0305004100046144}{\emph{Proc.~Camb.~Phil.~Soc.}
  {\bf 68} (1970) 345}.

\bibitem{edgarhoglund2002}
S.~B. Edgar and A.~H{\"o}glund, \emph{{Dimensionally dependent tensor
  identities by double antisymmetrization}},
  \href{http://dx.doi.org/10.1063/1.1425428}{\emph{J.~Math.~Phys.} {\bf 43}
  (2002) 659}, [\href{http://arxiv.org/abs/gr-qc/0105066}{{\tt
  gr-qc/0105066}}].

\bibitem{dappiaggihackpinamonti2009}
C.~Dappiaggi, T.-P. Hack and N.~Pinamonti, \emph{{The extended algebra of
  observables for Dirac fields and the trace anomaly of their stress-energy
  tensor}},
  \href{http://dx.doi.org/10.1142/S0129055X09003864}{\emph{Rev.~Math.~Phys.}
  {\bf 21} (2009) 1241}, [\href{http://arxiv.org/abs/0904.0612}{{\tt
  0904.0612}}].

\bibitem{christensenduff1979}
S.~M. Christensen and M.~J. Duff, \emph{{New gravitational index theorems and
  super theorems}},
  \href{http://dx.doi.org/10.1016/0550-3213(79)90516-9}{\emph{Nucl.~Phys.~B}
  {\bf 154} (1979) 301}.

\bibitem{mehta1990}
M.~R. Mehta, \emph{{Euclidean continuation of the Dirac fermion}},
  \href{http://dx.doi.org/10.1103/PhysRevLett.65.1983}{\emph{Phys.~Rev.~Lett.}
  {\bf 65} (1990) 1983}. [Erratum:
  \href{http://dx.doi.org/10.1103/PhysRevLett.66.522.2}{Phys.~Rev.~Lett.
  \textbf{66} (1991) 522}].

\bibitem{vannieuwenhuizenwaldron1996}
P.~van Nieuwenhuizen and A.~Waldron, \emph{{On Euclidean spinors and Wick
  rotations}},
  \href{http://dx.doi.org/10.1016/S0370-2693(96)01251-8}{\emph{Phys.~Lett.~B}
  {\bf 389} (1996) 29}, [\href{http://arxiv.org/abs/hep-th/9608174}{{\tt
  hep-th/9608174}}].

\bibitem{wetterich2011}
C.~Wetterich, \emph{{Spinors in euclidean field theory, complex structures and
  discrete symmetries}},
  \href{http://dx.doi.org/10.1016/j.nuclphysb.2011.06.013}{\emph{Nucl.~Phys.~B}
  {\bf 852} (2011) 174}, [\href{http://arxiv.org/abs/1002.3556}{{\tt
  1002.3556}}].

\bibitem{kupschthacker1990}
J.~Kupsch and W.~D. Thacker, \emph{{Euclidean Majorana and Weyl Spinors}},
  \href{http://dx.doi.org/10.1002/prop.2190380103}{\emph{Fortsch.~Phys.} {\bf
  38} (1990) 35}.

\bibitem{breitenlohnermaison1977}
P.~Breitenlohner and D.~Maison, \emph{{Dimensional renormalization and the
  action principle}},
  \href{http://dx.doi.org/10.1007/BF01609069}{\emph{Commun.~Math.~Phys.} {\bf
  52} (1977) 11}.

\bibitem{thooftveltman1972}
G.~'t~Hooft and M.~J.~G. Veltman, \emph{{Regularization and renormalization of
  gauge fields}},
  \href{http://dx.doi.org/10.1016/0550-3213(72)90279-9}{\emph{Nucl.~Phys.~B}
  {\bf 44} (1972) 189}.

\bibitem{kreimer1990}
D.~Kreimer, \emph{{The $\gamma_5$-problem and anomalies --- A Clifford algebra
  approach}},
  \href{http://dx.doi.org/10.1016/0370-2693(90)90461-E}{\emph{Phys.~Lett.~B}
  {\bf 237} (1990) 59}.

\bibitem{koernerkreimerschilcher1992}
J.~G. K{\"o}rner, D.~Kreimer and K.~Schilcher, \emph{{A practicable
  $\gamma_5$-scheme in dimensional regularization}},
  \href{http://dx.doi.org/10.1007/BF01559471}{\emph{Z.~Phys.~C} {\bf 54} (1992)
  503}.

\bibitem{thompsonhu1985}
G.~Thompson and H.-L. Yu, \emph{{$\gamma_5$ in dimensional regularization}},
  \href{http://dx.doi.org/10.1016/0370-2693(85)91397-8}{\emph{Phys.~Lett.~B}
  {\bf 151} (1985) 119}.

\bibitem{bonneau1980a}
G.~Bonneau, \emph{{Trace and axial anomalies in dimensional renormalization
  through Zimmermann-like identities}},
  \href{http://dx.doi.org/10.1016/0550-3213(80)90382-X}{\emph{Nucl.~Phys.~B}
  {\bf 171} (1980) 477}.

\bibitem{bonneau1980b}
G.~Bonneau, \emph{{Consistency in dimensional regularization with $\gamma_5$}},
  \href{http://dx.doi.org/10.1016/0370-2693(80)90232-4}{\emph{Phys.~Lett.~B}
  {\bf 96} (1980) 147}.

\bibitem{bonneau1981}
G.~Bonneau, \emph{{Preserving canonical Ward identities in dimensional
  regularization with a non-anticommuting $\gamma_5$}},
  \href{http://dx.doi.org/10.1016/0550-3213(81)90185-1}{\emph{Nucl.~Phys.~B}
  {\bf 177} (1981) 523}.

\bibitem{baikovilyin1991}
P.~A. Baikov and V.~A. Il'in, \emph{{Status of $\gamma^5$ in dimensional
  regularization}},
  \href{http://dx.doi.org/10.1007/BF01019107}{\emph{Theor.~Math.~Phys.} {\bf
  88} (1991) 789}.

\bibitem{bonoraetal2018b}
L.~Bonora, M.~Cvitan, P.~Dominis~Prester, S.~Giaccari, M.~Pauli{\v s}i{\'c} and
  T.~{\v S}temberga, \emph{{Axial gravity: a non-perturbative approach to split
  anomalies}},
  \href{http://dx.doi.org/10.1140/epjc/s10052-018-6141-1}{\emph{Eur.~Phys.~J.~C}
  {\bf 78} (2018) 652}, [\href{http://arxiv.org/abs/1807.01249}{{\tt
  1807.01249}}].

\bibitem{bastianellibroccoli2019}
F.~Bastianelli and M.~Broccoli, \emph{{On the trace anomaly of a Weyl fermion
  in a gauge background}},
  \href{http://dx.doi.org/10.1140/epjc/s10052-019-6799-z}{\emph{Eur.~Phys.~J.~C}
  {\bf 79} (2019) 292}, [\href{http://arxiv.org/abs/1808.03489}{{\tt
  1808.03489}}].

\bibitem{dreinerhabermartin2010}
H.~K. Dreiner, H.~E. Haber and S.~P. Martin, \emph{{Two-component spinor
  techniques and Feynman rules for quantum field theory and supersymmetry}},
  \href{http://dx.doi.org/10.1016/j.physrep.2010.05.002}{\emph{Phys.~Rept.}
  {\bf 494} (2010) 1}, [\href{http://arxiv.org/abs/0812.1594}{{\tt
  0812.1594}}].

\end{thebibliography}\endgroup

\end{document}